\documentclass[fleqn,usenatbib]{mnras}

\usepackage{newtxtext,newtxmath}

\usepackage[T1]{fontenc}

\DeclareRobustCommand{\VAN}[3]{#2}
\let\VANthebibliography\thebibliography
\def\thebibliography{\DeclareRobustCommand{\VAN}[3]{##3}\VANthebibliography}

\usepackage{graphicx}	
\usepackage{amsmath}	
\usepackage{float}
\usepackage[caption = false]{subfig}
\usepackage{longtable}
\usepackage{nicefrac}
\usepackage{xltabular}
\usepackage{tabularx}
\usepackage{rotating}
\usepackage{adjustbox}
\usepackage{lscape}
\usepackage{comment}
\usepackage{ragged2e}
\usepackage{hyperref}
\usepackage{geometry}
\usepackage{enumitem}
\usepackage{blindtext}
\usepackage{marvosym}

\newcommand\clearrow{\global\let\rowmac\relax}
\clearrow

\def\I{{\em INTEGRAL}}
\def\M{{\em MAXI}} 
\def\R{{\em RXTE}} 
\def\gcm2{\,g~cm$^{-2}$}
 
\def\ergcs{\,erg~cm$^{-2}$~s$^{-1}$} 
\def\ergcm{\,erg~cm$^{-2}$} 
\def\ergps{\,erg~s$^{-1}$}

\date{Accepted 2023 January 31. Received 2023 January 31; in original form 2021 June 9}

\pubyear{2023}

\begin{document}
\title[Long-burst catalogue]
{A catalog of unusually long thermonuclear bursts on neutron stars}

\author[Alizai et al.]{K.~Alizai,$^{1}$
J.~Chenevez,$^{1}$\thanks{E-mail: jerome@space.dtu.dk}
A.~Cumming,$^{2}$
N.~Degenaar,$^{3}$
M.~Falanga,$^{4}$
D.~K.~Galloway,$^{5}$
\newauthor
J.~J.~M.~in~'t~Zand,$^{6}$
G. K.~Jaisawal,$^{1}$
L.~Keek,$^{7}$
E.~Kuulkers,$^{8}$
N.~Lampe,$^{5}$
H.~Schatz,$^{9}$
\newauthor
M.~Serino$^{10}$
\\
$^1$National Space Institute, Technical University of Denmark,
Elektrovej 327, 2800 Kongens Lyngby, Denmark \\
$^2$Physics Dept., McGill University, 3600 Rue University, Montreal, QC, H3A 2T8, Canada\\
$^3$Anton Pannekoek Institute for Astronomy, University of Amsterdam,
Postbus 94249, NL-1090 GE Amsterdam, the Netherlands\\
$^4$International Space Science Institute, Hallerstrasse 6, 3012 Bern, Switzerland\\
$^5$School of Physics \& Astronomy, Monash University, VIC 3800, Australia \\
$^6$SRON Netherlands Institute for Space Research, 
Niels Bohrweg 4, 2333 CA Leiden, The Netherlands  \\
$^7$Cosine measurement systems, Oosteinde 36, Warmond, 2361 HE, The Netherlands \\
$^8$ESA/ESTEC, Keplerlaan 1, 2201 AZ Noordwijk, The Netherlands \\
$^9$Dept. of Physics and Astronomy and National Superconducting Cyclotron Laboratory, Michigan State University, East Lansing, MI 48824\\
$^{10}$Department of Physics and Mathematics, Aoyama Gakuin University,
5‐10‐1 Fuchinobe, Chuo‐ku, Sagamihara, Kanagawa 252‐5258, Japan\\
} 
 
\maketitle


\begin{abstract} 
Rare, energetic (long) thermonuclear (Type I) X-ray bursts are classified either as intermediate-duration or ``super’' bursts, based on their duration. Intermediate-duration bursts lasting a few to tens of minutes are thought to arise from the thermonuclear runaway of a relatively thick ($\approx 10^{10}$ g cm$^{-2}$) helium layer, while superbursts lasting hours are attributed to the detonation of an underlying carbon layer. 
We present a catalogue of 84 long thermonuclear bursts from 40 low-mass X-ray binaries, and defined from a new set of criteria distinguishing them from the more frequent short bursts. 
The three criteria are: 1) a total energy release longer than $10^{40}$~ergs, 2) a photospheric radius expansion phase longer than $10$~s, and 3) a burst timescale longer than $70$~s.
This work is based on a comprehensive systematic analysis of 70 bursts found with \textit{INTEGRAL}, \textit{RXTE}, \textit{Swift}, \textit{BeppoSAX}, \textit{MAXI} and \textit{NICER}, as well as 14 long bursts from the literature that were detected with earlier generations of X-ray instruments. For each burst, we measure its peak flux and fluence, which eventually allows us to confirm the
distinction between intermediate-duration bursts and superbursts. Additionally, we list 18 bursts 
that only partially meet the above inclusion criteria, possibly bridging the gap between normal and intermediate-duration bursts.
With this catalogue, we significantly increase the number of long-duration bursts included in the MINBAR and thereby provide a substantial sample of these rare X-ray bursts for further study.
\end{abstract}



\newpage
\section{Introduction}
\label{sec:Introduction} 
Type-I X-ray bursts have been discovered more than five decades ago (\cite{Belian}; \cite{Belian76}; \cite{Grindlay76}) and identified as due to the unstable thermonuclear ignition of hydrogen and/or helium on the surface of neutron stars (\cite{Han_Horn75}; \cite{MarCav}; \cite{woos76}; \cite{Joss},; \cite{lamb78}; \cite{taam}).
The thermonuclear fuel is accreted from a companion star through Roche-lobe overflow in a low-mass X-ray binary (LMXB) system. Stable burning of hydrogen may create additional helium below the hydrogen layer on the neutron star surface. When the pressure and temperature at the base of the accumulated H/He reach ignition conditions, explosive thermonuclear burning leads to an X-ray burst. The burning of the accumulated matter can produce heavy elements well above the iron group if sufficient hydrogen remains in the envelope at the time of ignition. The duration of the burst can vary between 10 seconds and tens of hours, depending on the ignition depth and the metallicity of the fuel layer (see \cite{Lew93}; \cite{StrohBil06}, and \cite{Gal_keek17} and references therein). The effective temperature of the photosphere during the burst can exceed 30 MK, while the luminosity at the peak of the burst can reach the local Eddington limit where the radiation pressure overcomes the gravitational pull, causing the photosphere to expand. The Eddington luminosity reached during the photospheric radius expansion (PRE) phase depends on the atmospheric composition (for a neutron star mass of 1.4 M$_\odot$, $L_{\rm Edd}\approx 2.1 \times 10^{38}$ \ergps for an atmosphere with cosmic abundances and $L_{\rm Edd}\approx 3.5 \times 10^{38}$ \ergps for an He-rich atmosphere; see, e.g., \cite{Lew93}). Moreover, it can be used to estimate the distance to the bursting sources, assuming a given atmosphere composition at the time of the burst.

X-ray binaries that have shown Type-I X-ray bursts can be divided into two sub-groups. The first group consists of a low-mass star as the donor and the neutron star as the accretor. As the donor star has to fit in the binary, these systems have orbital periods longer than 80 min. The second group of systems is called ultra-compact X-ray binaries (UCXB) (\cite{nelson86}; \cite{nelemans}). These systems are characterized by their short orbital periods ($< 80$ min) that can 
only fit a degenerate companion star. The hydrogen-poor environment of UCXBs results in mainly pure helium flashes that reach the Eddington luminosity and produce PRE bursts. 

The most common X-ray bursts ($\approx 99\%$ of all bursts observed) have a duration of up to one minute and are caused by unstable burning of accreted He or mixed H/He. 
The burning that can occur during a burst  
depends on the temperature of the fuel envelope before ignition. This is set by the accretion rate, the stable H burning that heats the envelope, as well as pycnonuclear reactions and electron captures in the neutron-star crust (e.g. \cite{Fuji81} \& \cite{Gal_keek17}).
At low accretion rates ($< 0.01 \dot{M}_{\rm Edd}$), the stable burning of hydrogen, via the hot CNO cycle, continues for a long enough time so that all the hydrogen is consumed before the burst. This results in the unstable burning of a pure helium layer (\cite{Peng}). At higher accretion rates $> 0.1 \dot{M}_{\rm Edd}$, time is not sufficient to burn the hydrogen completely before ignition, which results in a burst with mixed H/He composition. 
The largest observation sample of these "common" bursts is assembled in the Multi-INstrument Burst ARchive \citep[MINBAR;][]{Gal20}.

In fact, the first burst ever detected \citep[e.g.;][from Cen X-4 in 1969]{Belian, Kuul09} belongs to a population that is distinct from the “common” bursts in being significantly longer (about 10 minutes). A dozen other long bursts were detected with the first generation of X-ray instruments during the last three decades of the $20^{th}$ century. 
In the past two decades, observations of such rare long bursts have increased significantly, thanks to the wide field of view (FOV) monitoring capabilities of {\em BeppoSAX}, \I, \M, \R, and {\em Swift} 
\citep[e.g.;][]{intZ07, Chenevez08, Serino16, Cor00, intZ19, alizai}.
Long type-I bursts are of interest because their ignition conditions are sensitive to the crust composition and temperature of the neutron star (\cite{C06}). This also makes long type-I X-ray bursts naturally more rare, as it takes longer time to accumulate the required fuel to achieve ignition.
These bursts can be subdivided into two groups: \textit{intermediate-duration bursts} and \textit{superbursts}.
Intermediate-duration bursts have been observed to last from several minutes up to one hour, for a total energy release up to a few times $10^{41}$ ergs. They are thought to arise from the ignition of helium at column depths of $\approx 10^{10}$\gcm2 (\cite{intZ05}; \cite{C06}).
Intermediate-duration bursts can furthermore be prime candidates for improving our understanding of low-level accretion physics \citep[e.g.;][]{intZ11, intZ12, degenaar18}.
A deep layer of helium can be accumulated indirectly in sources with a hydrogen-rich donor star, or directly in sources with a helium-rich optical companion. In the former, the thick layer of helium is accumulated via hydrogen burning at shallow depths either stably or in weak flashes (\cite{Peng}; \cite{Coop}). 
In the latter, the helium is accumulated at very low accretion rates of $\approx 0.001 - 0.01 \dot{M}_{\rm Edd}$. In either case, when ignited, this large pile of helium results in an intermediate-duration burst.

Superbursts last from hours to a whole day and may release $\sim10^{42}$ ergs, i.e. $\approx1000$ times more energy than common bursts. They are thought to be triggered by the burning of a deep layer of carbon, with possible additional energy release from photodisintegration of heavy elements produced in previous rp-process burning episodes. For deep ignition of carbon, neutrino cooling can be an important mechanism of energy loss in addition to cooling from the neutron star surface \citep[e.g.;][]{CB01, Strohbrown02, C06}. Numerical models have shown that carbon fractions of at least $10-20\%$ in the thermonuclear fuel are necessary to explain the observed superbursts (\cite{CB01}; \cite{Strohbrown02}; \cite{C06}). However, an additional, unidentified, shallow heating source from the crust is required to ignite carbon at the measured depths \citep[e.g.;][]{C06, Keek08, deibel16}.
X-ray sources that have shown superbursts have typical accretion rates around $0.1 \dot{M}_{\rm Edd}$ (where $\dot{M}_{\rm Edd} = 2\times10^{-8}$ M$_\odot$/yr), though superbursts have also been observed at extreme values of accretion rates (e.g. 4U~0614$+$09 at $\approx 0.01 \dot{M}_{\rm Edd}$ (\cite{Kuul10}) and GX~17$+$2 at $\approx 1 \dot{M}_{\rm Edd}$ (\cite{intZ04})). 
At the time of writing, 28 superbursts have been reported from 16 sources (see list in \cite{intZ17b})\footnote{One more superburst is known (seen by MAXI from 4U~1608-52 on July 16, 2020) that has not been reported yet (Iwakiri et al., in prep.).}.

\begin{figure}
\centering
\subfloat{\includegraphics[width = 2.2in,angle=-90]{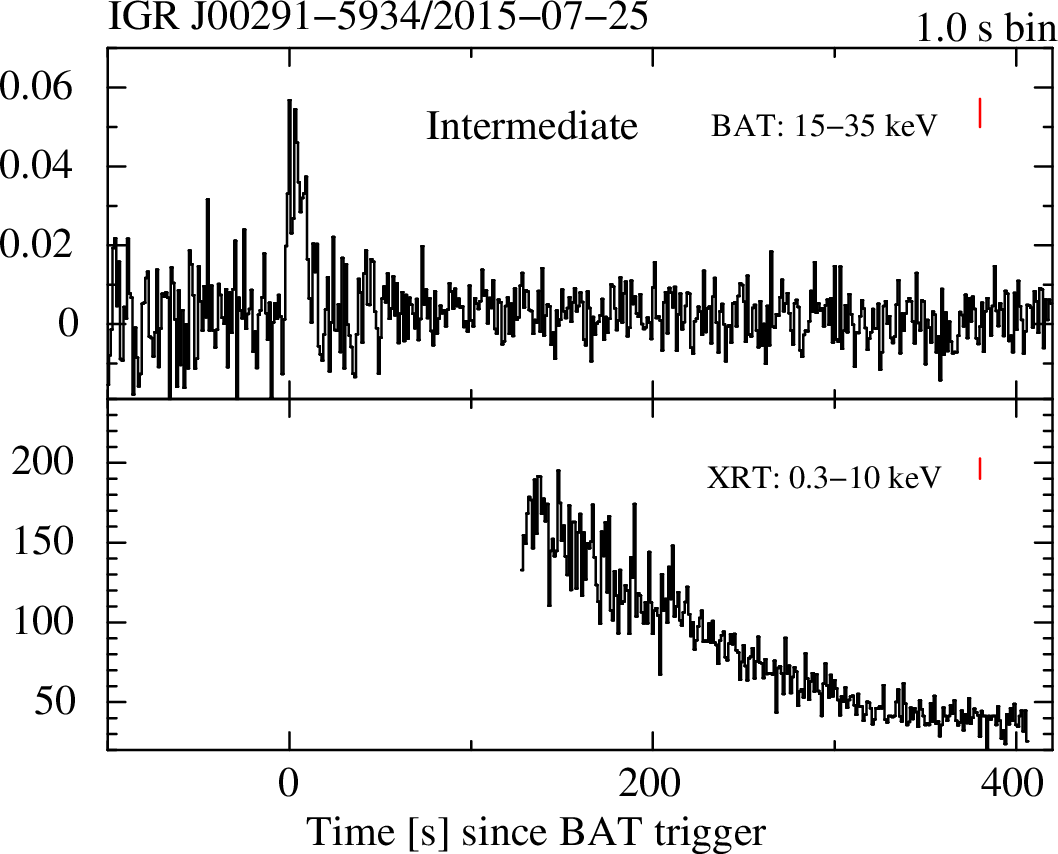}}\\  
\subfloat{\includegraphics[width = 2.2in,angle=-90]{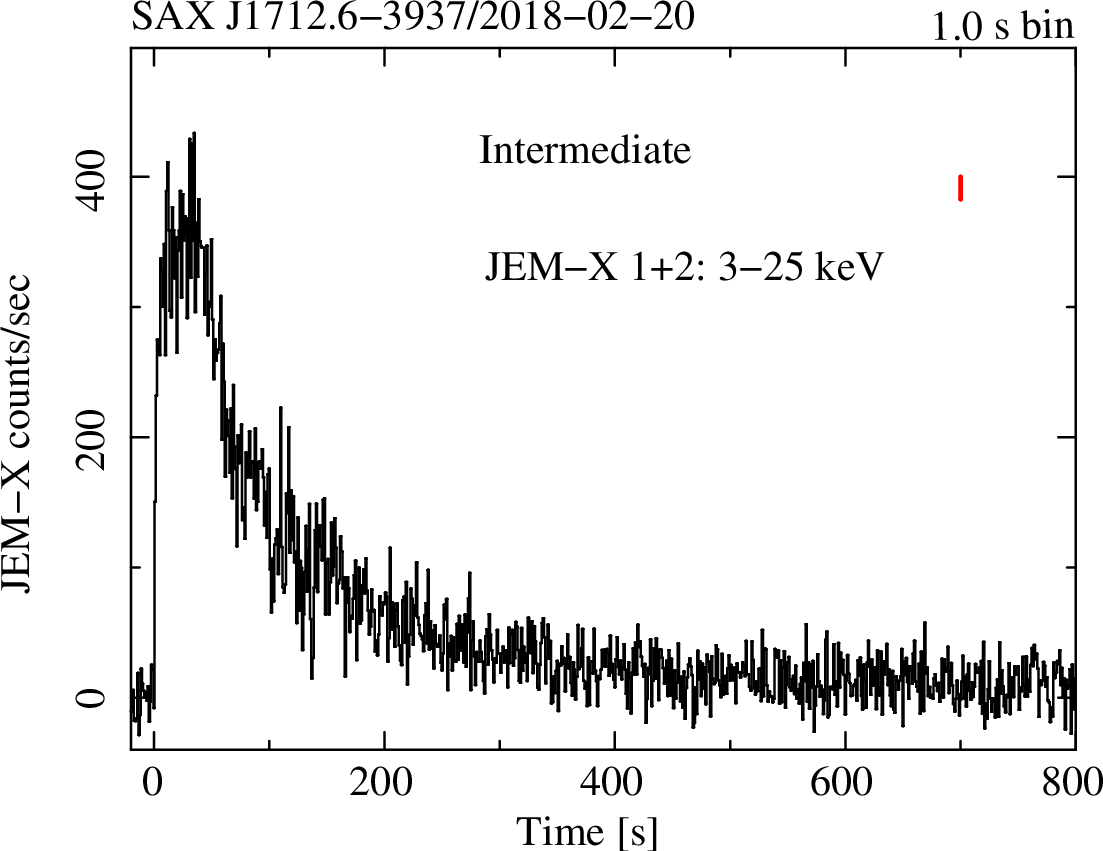}}\\
\subfloat{\includegraphics[width = 2.3in,angle=-90]{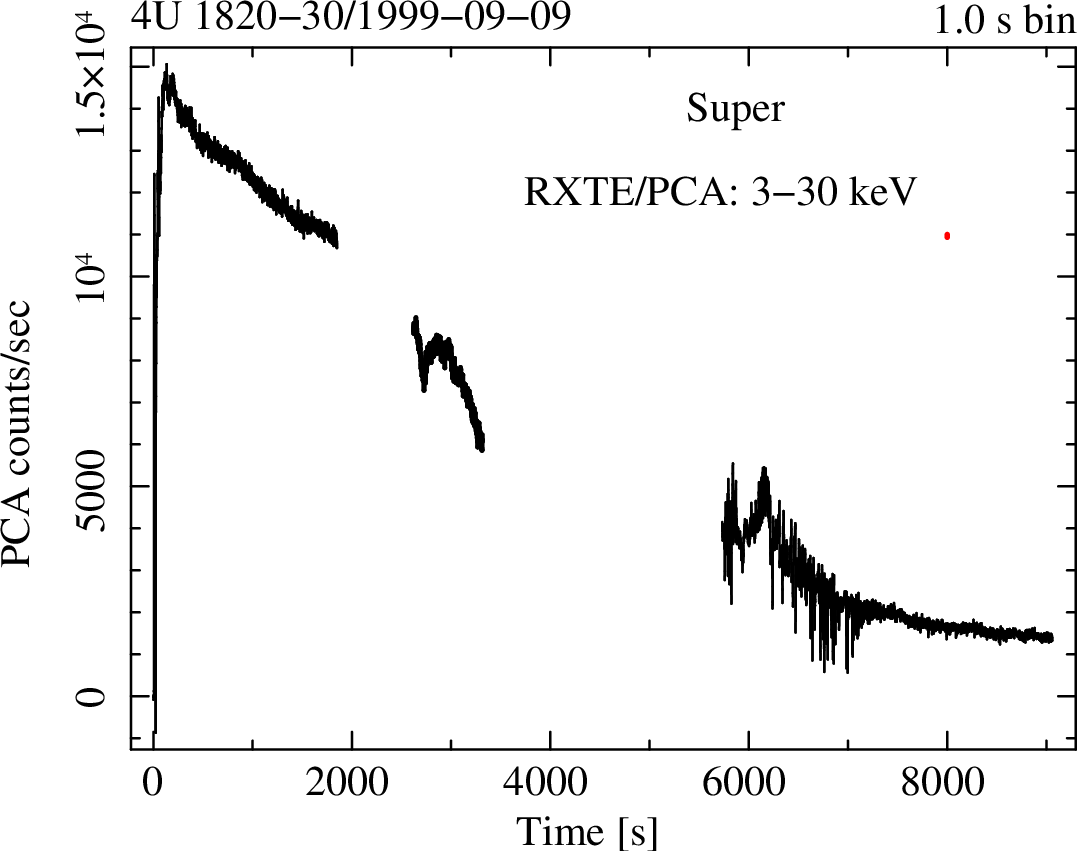}}        
\caption{Examples of the diversity of long bursts observed with three different instruments. \textit{Top}: The intermediate-duration burst detected by {\em Swift}, from (the accretion-powered pulsar) IGR~J00291$-$5934 (\protect\cite{defalco17a}). \textit{Middle}: An intermediate-duration burst from SAX~J1712.6-3739 observed with \protect\I/JEM-X (\protect\cite{alizai}). A superburst from 4U~1820-30 observed with RXTE/PCA (\protect\cite{Strohbrown02}). The pre-burst count rate has been subtracted for all light curves. The red markers on each plot indicate the typical uncertainty on each (light curve) data point.}
\label{fig:1}
\end{figure}
 
Here, we present a comprehensive catalogue of long X-ray bursts based on systematic light curve and spectral analyses of data from satellites that have detected multiple long bursts. We do not re-analyze data from a few old missions due to their data acquisition and/or quality issues. 
Examples of different long bursts from our {catalogue} are shown in Figure \ref{fig:1}. 
Table \ref{tab:overview} displays the characteristics of the 40 known X-ray sources that have shown at least one long burst.
The present work supplements the MINBAR sample with a uniform analysis of all long bursts observed with a large diversity of X-ray instruments and with superbursts not included in MINBAR.

Section \ref{sec:instruments} describes the instruments whose data are included in our analyses,  
which are presented in
section \ref{sec:data_analysis}. Our observational results are gathered in the catalog presented in section \ref{sec:catalogue}.  
In Section \ref{sec:discussion}, we discuss the main outcomes of the present study and in  
section \ref{sec:conclusion}, we present our conclusions.


\section{Instrument descriptions}
\label{sec:instruments}
In this section, we describe the general characteristics and data reduction procedures of the following instruments (the number of long bursts observed with each instrument is indicated in the parentheses): \I/JEM-X \& ISGRI (13), \R/PCA, and ASM (18), \textit{MAXI} (11), \textit{Swift}/BAT \& XRT (12), and \textit{BeppoSAX}/WFC (13).  Three more long bursts were observed (almost) simultaneously by two instruments: \textit{MAXI + NICER} (1), JEM-X + \textit{MAXI} (1) and \textit{MAXI} + XRT (1). Thanks to the large field of view of most of these instruments, they have been successful in detecting rare events, enabling serendipitous observations of long bursts. Whenever we quote spectral or spatial resolutions or widths in the remainder of this paper, we refer to the value of the full width at half maximum (FWHM).


\subsection{INTEGRAL}
\label{subsec:INTEGRAL}
\textit{INTEGRAL} was launched on 17 October 2002 and is still operational. The Joint European X-ray Monitor (JEM-X) is the soft X-ray instrument onboard the {\textit{INTEGRAL}} satellite, with an energy range of $3 - 35$~keV. JEM-X consists of two identical co-aligned coded-mask telescopes, each with a high-pressure imaging microstrip Xenon gas detector, a $4.8^{\circ}$ diameter FoV, $3.3$~arcmin angular resolution, and a spectral resolution of 1.2~keV at 10~keV (\cite{lund03}).  
The Integral Soft Gamma-Ray Imager (ISGRI) is the top layer of the Imager on Board Integral Satellite (\textit{IBIS}). ISGRI has an energy range of $15 - 1000$~keV, with a diminishing sensitivity below $20$~keV, making it only able to detect the very high-energy tail of the X-ray burst spectra, an $8.3^{\circ} \times 8.0^{\circ}$ FOV (fully coded), 12~arcmin angular resolution, and a spectral resolution of 8~keV at 100~keV (\cite{u03}).

A typical observation with the \textit{INTEGRAL} satellite consists of multiple pointings, referred to as science windows (ScW), separated by $\simeq 2^{\circ}$. A typical ScW lasts between 1800 s to 3600 s (\cite{Jensen}).
JEM-X has detected sixteen out of the seventeen long X-ray bursts observed by \textit{INTEGRAL}. Only three had sufficiently hard photons that IBIS/ISGRI detected from the sixteen bursts detected by JEM-X. Only one burst, from SLX~1735-269, has been detected by ISGRI alone (\cite{sguera07}; \cite{alizai}). 
The data reduction of \textit{INTEGRAL} satellite was performed with the Offline Science Analysis software (\textit{OSA}), version 11.


\subsection{Rossi X-ray Timing Explorer}
\label{subsec:RXTE}
The \textit{Rossi X-ray Timing Explorer} (\R) was launched on 30 December 1995 and decommissioned on 3 January 2012. It carried three instruments of which the Proportional Counter Array (PCA) and the All-Sky Monitor (ASM) are relevant for our purposes. 


\subsubsection{Proportional Counter Array}
\label{subsubsec:PCA}
The PCA (\cite{Jahoda}) consists of five identical proportional counter units (PCU's), sensitive in the energy range 2 - 60 keV, a geometric photon-collecting area of 8000 cm$^{2}$, a spectral resolution of about 18\% at 6 keV. Each PCU has a collimator admitting photons within a 1$^{\circ}$ radius of the pointing direction. The PCA has 2 Standard data modes; Standard mode 1 provides a time resolution of $0.125$ s but no spectral information. Standard mode 2 provides high spectral resolution data (129 channels over the PCA energy band) every $16$ s. Simultaneously with the two Standard modes, data is recorded in user-selected higher time-resolution modes, with lower spectral readout resolution (B-modes or E-modes). 
For this study, we use the E-mode and the two Standard modes data to produce high time resolution light curves and spectra (using the HEASARC tasks \texttt{seextrct 4.2e} and \texttt{saextrct 4.3a}. The bursts detected by PCA are bright enough (more than $1000$ c/s at the peak) for deadtime to be an issue. We have therefore corrected all the PCA data for deadtime in this work\footnote{\url{https://heasarc.gsfc.nasa.gov/docs/xte/recipes/pca_deadtime.html}}. Except for the superbursts, all long bursts detected with the PCA are included in MINBAR (\cite{Gal20}). The count rates of the PCA light curves presented in this paper are summed over the active PCUs.


\subsubsection{All-Sky Monitor}
\label{subsubsec:ASM}
The All-Sky Monitor (ASM: \cite{levine96}) consists of 3 Scanning Shadow Cameras (SSC), with each having a position-sensitive proportional counter and a 6$^{\circ}$ $\times$ 90$^{\circ}$ FOV. ASM is sensitive in the energy range 1.3 – 12.1~keV and covers 80$\%$ of the sky every 90 minutes. The ASM data is accumulated in 90 seconds dwells, after which the orientation of SSCs is changed. The data products, consisting of light curves and time series of intensities in three energy bands, can be downloaded as FITS files from the ASM data products page \footnote{\url{https://heasarc.gsfc.nasa.gov/docs/xte/asm_products.html}}.

\subsection{\textit{Neil Gehrels Swift} observatory}
\label{subsec:Swift}
The \textit{Neil Gehrels Swift} observatory was launched on 20 November 2004. It is a dedicated mission for searching and studying gamma-ray bursts (GRBs: \cite{Gehrels04}) but has also detected multiple long X-ray bursts (\cite{intZ19}). \textit{Swift} carries three instruments, but only two are relevant for our purposes: the Burst Alert Telescope (BAT) and the X-ray Telescope (XRT).  
For all long bursts in our study from the XRT, only the tail is detected.


\subsubsection{Burst Alert Telescope}
\label{subsubsec:BAT}
The BAT is a coded aperture camera with a FOV of $120^{\circ} \times 60^{\circ}$, accounting for $18\%$ of the sky, an angular resolution of $17$ arcmin (on-axis), and localization accuracy of $1-4$ arcmin. The telescope has a CdZnTe detector array, giving it a bandpass of $15-150$~keV with an energy resolution of 5~keV at 60~keV (\cite{Barth}).
We use the HEASARC task \textit{batbinevt}\footnote{\url{https://heasarc.gsfc.nasa.gov/ftools/caldb/help/batbinevt.html}} (HEASOFT 6.22.1) to extract light curves and spectra from BAT event data. Light curves can be created in the energy bands between $15$~keV and $150$~keV for triggered data. The BAT spectra can be extracted for a whole observation or over a specific time interval. 


\subsubsection{X-ray Telescope}
\label{subsubsec:XRT}
When the BAT has located a burst that meets the triggering threshold (typically $8 \sigma$ above the background noise level), the spacecraft slews autonomously to that direction and brings the source into the 23.6 $\times$ 23.6 arcmin FoV of the X-ray Telescope (XRT). The XRT (\cite{Burrows}) has an energy range of $0.2$~keV to $10$~keV with an angular resolution of 18-22~arcsec and a position accuracy of $3$~arcsec. The detector is a $600 \times 600$ pixels CCD camera  with a spectral resolution of $140$ eV at $5.9$~keV, an effective area of $125$~cm$^{2}$ at $1.5$~keV and $20$~cm$^{2}$ at $8$~keV. All the bursts from \textit{Swift} reported in this study involve automatic slews.
We use the online facility \textit{Build Swift-XRT products}\footnote{\url{https://www.swift.ac.uk/user_objects/}} to produce light curves and spectra in Windowed Timing (WT) mode, which has $1.8$ ms time resolution and is useful for fluxes below $5000$ mCrabs (\cite{evans07} \& \cite{evans09}).


\subsection{BeppoSAX Wide Field Camera (WFC)}
\label{subsec:WFC}
The Italian-Dutch satellite \textit{BeppoSAX} was launched in April 1996 and operated from June 1996 to April 2002 (\cite{boella}). Onboard were two identical Wide Field Cameras (WFCs; \cite{Jager}), with $40^{\circ} \times 40^{\circ}$ FoV, an energy resolution of 1.1 keV at 6~keV, and $5$ arcmin angular resolution in the energy range of $2-25$~keV. \textit{BeppoSAX}/WFC has observed a total of $13$ long bursts.    


\subsection{Monitor of All-sky X-ray Image (MAXI)}
\label{subsec:MAXI}
The Monitor of All-sky X-ray Image (\textit{MAXI}) was launched in April 2009 and is still active. One of the main scientific instruments of \textit{MAXI} (\cite{matsuoka})  onboard the International Space Station (ISS) is the Gas Slit Camera (GSC; \cite{Mihara}). GSC observes $ \approx 85\%$ of the whole sky every $92$ minutes in the $2-20$~keV band, with an energy resolution of 1.1~keV at 5.9~keV, and a combined geometrical collecting area of $\approx$ 5350~cm$^{2}$. Sources showing long bursts are observed for $40$ - $120$ s (corresponding to $1$ - $3$ transit scans (\cite{Sugi11}). In this study, we report on $11$ bursts observed only by \textit{MAXI}/GCS (\cite{iwak1}; \cite{Serino16}), and additionally $3$ bursts that are also observed by other observatories (\cite{Serino16}; \cite{intZ17b}). We also use the \textit{MAXI}/GSC long-term light curves of sources that have shown long bursts.  


\section{Data Analysis}
\label{sec:data_analysis}

In this section we explain how we have selected our input data, and present the systematic analysis methods used to derive the burst properties included in our catalogue. The light curve analysis method is instrument specific, while the spectral analysis method is the same across all the instruments. 
We use the X-ray spectral fitting package \texttt{XSPEC} (\cite{Arnaud96}) to perform the spectroscopy for this study.
All uncertainties are given at a $1 \sigma$ confidence level.


\subsection{Light curve analysis}
\label{subsec:lightcurve_ana}
The bolometric peak flux of bursts observed with \textit{INTEGRAL}/JEM-X and \textit{BeppoSAX}/WFC is derived using the following equation:

\begin{equation} \label{eq1}
\begin{split}
 F_{\rm peak}^{\rm bol} &= \frac{\overline{F}_{\rm bol}}{\overline{F}_{\rm cts}} \times F_{\rm cts}^{\rm peak}\\
\end{split}
\end{equation}
where $F_{\rm cts}^{\rm peak}$ is the background-subtracted burst peak count-rate from the instrument $1$~s-bin light curve; $\overline{F}_{\rm bol}$ is the bolometric flux, obtained from spectral analysis (see below) over the time interval surrounding the burst peak; $\overline{F}_{\rm cts}$ is the average count-rate over the same time interval.

For sources that have constraints on the inclination angle of the accretion disk, we correct the burst peak fluxes for anisotropy with the factor $\xi_{\rm b}$, adopted from \cite{He16}, but
neglecting the reflection contribution from the accretion disk, which leads to: 
\begin{equation} \label{eq2}
    \xi_{\rm b} =  \frac{2}{\cos{i}+1}
\end{equation}

The bolometric peak flux of bursts observed with \textit{RXTE}/PCA is derived through spectroscopy (see Section \ref{subsec:spectral_analysis}) of a 1~s spectrum extracted from the peak of the burst, localized using the instrument light curve with 1~s bin size. Whenever a burst is observed in the Standard 2 data mode only, we use equation \ref{eq1} to obtain the bolometric peak flux. In that case, the $\overline{F}_{\rm bol}$ is derived from the 16~s spectrum at the peak, and the $F_{\rm cts}^{\rm peak}$ from the Standard 1 light curve with 1~s bin size.

For \textit{Swift} bursts the onset is only available in the hard X-ray region ($15-35$~keV), where the intensity of the bursts is relatively low. The e-folding times quoted are obtained by fitting the tail of the bursts detected in XRT. 
The bolometric peak flux is estimated by applying Eq. \ref{eq1}. We assume a constant bolometric flux during the PRE phase, as the peak in the BAT count rate coincides with the end of the PRE phase. Using BAT count rates to calculate the peak flux results in a larger error. 

Due to the duty cycle of \textit{MAXI} on a given source, we can only provide a lower limit for the peak flux of bursts detetcted by this instrument. Furthermore, we can only fit the light curves of bursts that are detected in at least 3 orbits, providing three data points, and quote decay times for them. We fit the \textit{MAXI}/GSC light curve in the energy band $2 - 20$~keV. We also retrieve the \textit{MAXI}/GSC long-term monitoring light curves for the sources in our sample, that are also included in the \textit{MAXI} source catalogue\footnote{\url{http://maxi.riken.jp/top/lc.html}}.   

The $90$-second dwells of \textit{RXTE}/ASM prevent us to measure the bolometric flux at the very peak. As for \textit{MAXI}/GSC, we also retrieve the \textit{RXTE}/ASM long-term monitoring light curves for the sources in our sample, that are also included in ASM long-term observed sources list\footnote{\url{http://xte.mit.edu/ASM_lc.html}}. 

\subsection{Spectral analysis}
\label{subsec:spectral_analysis}

\subsubsection{Persistent emission}
\label{subsubsec:persistent_spectral}
The observed signal during bursts generally consists of two main components; 1) the emission produced by the burning on the neutron star surface, and 2) the persistent emission originating in the accretion flow. As a consequence of the serendipitous nature of the detection (apart from \textit{Swift}/XRT observations initialized by BAT triggers), a significant number of observations lack the onset of the bursts. Any source flux immediately before the burst is also missing in these cases.
For time-resolved spectroscopy, we extract persistent flux before the burst, where this is available. In some other cases, we choose to extract persistent flux from the orbit/observation before the burst, but within 24~h. For sources with persistent flux levels below the source detection limit of the instrument in question, we ignore the persistent flux.

We model the persistent emission in order to obtain the bolometric persistent flux for the calculation of $\gamma$ (see eq. \ref{eq6} in \ref{subsubsec:burst_em}). The most common model we used is a simple power law (\texttt{PO} in \texttt{XSPEC}). The second most common model we used is also a power-law, but with a high energy roll-off (\texttt{CUTOFFPL} in \texttt{XSPEC}). In rare cases we have modelled the persistent emission with an accretion disk consisting of multiple black body components (\texttt{diskBB} in \texttt{XSPEC}) or a combination of the mentioned models. We also investigated the persistent emission evolution during bursts by applying the enhanced persistent emission method (see Section \ref{subsubsec:burst_em}). The persistent model parameters, for the bursts where this was modelled, are provided in the online catalogue.

\subsubsection{Burst flux and variable persistent flux during a burst}
\label{subsubsec:burst_em}
We investigate all spectral data by applying the same models within \texttt{XSPEC} when possible, so as to perform a systematic time-resolved spectral analysis of 70 bursts (excluding the bursts from the literature study).

We model the absorption column density with the \texttt{XSPEC}-model \texttt{TBABS} (\cite{W2000}). 
Due to the limited statistical quality of our data in general, we model the emission from the neutron star surface, during a burst, with a simple black body (\texttt{BBODYRAD} in \texttt{XSPEC}), as is customary for burst emission.

There are notable exceptions where the burst emission is not well-fitted by a black-body continuum, e.g. 4U~1820-30 (\cite{Strohbrown02}), 4U~1636-536 (\cite{Keek14a}), and IGR~J17062-6143 (\cite{degenaar13, Keek17}), that we present in our online catalogue. In this paper, we present our results using a simple black-body model for the burst emission.

To test improvements to our spectral fits and investigate the evolution of the persistent emission during a burst, we include the persistent spectral model to the fits in addition to a black body. We fix the parameters of the persistent spectral model to the pre-burst best-fit values and add a \texttt{constant} component, to represent a factor whose value may change at each time interval. This value, referred as $f_{a}$ from now on, indicates the variation of the persistent flux during a burst. This approach does improve the spectral fits of \textit{RXTE}/PCA bursts \citep[see also][]{wo13, Keek14a, wo15}, but due to the limited spectral quality of the rest of the data used for this study, the $f_{a}$ parameter is ill-constrained. The results of this approach are presented in the online catalogue.

Here, we present results obtained with the previously regarded "standard" spectral analysis method (\cite{sztajno}), where the pre-burst flux is subtracted as a background from the burst flux. This method is not correct when the photosphere significantly changes the persistent flux, making the late cooling phase especially bad for fitting (\cite{kuul02b}), the limited statistical quality of our data prevents us from applying more sophisticated analysis methods.

\subsection{Basic burst parameters} 
\label{subsubsec:burst_param}
For each burst in the catalogue, we derive the following parameters.

The measured net burst fluence, \textit{E$_{\rm obs}$} (in \ergcm), is the sum of the integrated bolometric fluxes obtained through time-resolved spectroscopy. For this parameter we refrain from interpolating over data gaps, and extrapolating to the start and end of a burst when either of these two times is not observed. Consequently, the net \textit{E$_{\rm obs}$} we measure may differ from previously-published values, and it is a lower limit to the actual fluence for bursts not totally covered by observations. The estimated uncertainty on this parameter is obtained by evaluating the uncertainties on flux values from the time-resolved spectral analysis.

Nevertheless, we also want to derive the burst total fluence, $E_{\rm b}$, so as to calculate the corresponding total energy, $E_{\rm tot}=4\pi d^2\times E_{\rm b}$, irradiated during a burst.
For bursts that are fully covered by observations, \textit{E$_{\rm b}$} is equal to \textit{E$_{\rm obs}$}. For bursts not entirely covered by observations, we fit the fragmented bolometric light curve with a power law of the form (\cite{intZ14a}):

\onecolumn
\begin{table}
\caption{List of 40 sources that have shown at least one long burst, adopted from \citep{Gal20}}
\begin{tabular}{l c c c c c c c c c}
\hline \hline
 {\textbf{Source}} & {\textbf{Type}$^{a}$} & {\textbf{R.A.}} & {\textbf{Dec.}} & {\textit{\textbf{i}}} & {\textit{\textbf{d}}} & {\textit{\textbf{P$_{\rm orb}$}}} & {\textbf{**I}} & \textbf{***S} & \textbf{Ref.}\\
  &  & \textbf{(Eq. 2000.0)} & \textbf{(Eq. 2000.0)} & {\textbf{($^{\circ}$)}} & {\textbf{(kpc)}} & {\textbf{(hr)}} \\
\hline \hline
IGR J00291-5934 & PT & $00^{h}29^{m}03^{s}.050$ & $+59^{\circ}34'18''.91$ & - & $4.2 \pm 0.5$* & $2.46$ & 1 & - & [1], [2]\\
4U 0614+091 & ACR & $06^{h}17^{m}07^{s}.35$ & $+09^{\circ}08'13''.4$ & $50.0 - 62.0$ & $2.59 \pm 0.03$ & $0.855$ & 1 & 2 & [3], [4], [5] \\
2S 0918-549 & C & $09^{h}20^{m}26^{s}.473$ & $-55^{\circ}12'24''.47$ & - & $3.9 \pm 0.2$ & $0.29$ & 2 & - & [6], [7], [8], [9]\\
4U 1246-588 & C\textbf{\textsuperscript{\Cross}} & $12^{h}49^{m}39^{s}.364$ & $-59^{\circ}05'14''.68$ & $0 - 70.0$ & $3.8 \pm 0.2$ & - & 1 & - & [10], [11], [12]\\
4U 1254-690 & D & $12^{h}57^{m}37^{s}.15$ & $-69^{\circ}17'21''.0$ & $68.0 - 73.0$ & $<7.9$ & $3.93$ & - & 1 & [13], [14], [15], [16]\\
Cen X-4 & RT & $14^{h}58^{m}21^{s}.92$ & $-31^{\circ}40'07''.4$ & - & $ \approx 1.2$ & $15.1$ & 1 & - & [17], [18], [19]\\
4U 1608-522 & AOT & $16^{h}12^{m}43^{s}.0$ & $-52^{\circ}25'23''$ & $25.0 - 50.0$ & $3.2 \pm 0.3$ & $12.9$ & - & 1 & [20], [21], [22]\\
4U 1636-536 & AOR & $16^{h}40^{m}55^{s}.57$ & $-53^{\circ}45'05''.2$ & $36.0 - 74.0$ & $5.0 \pm 0.5$ & $3.8$ & - & 4 & [23], [24], [25], [26]\\
XTE J1701-407  & T & $17^{h}01^{m}44^{s}.3$ & $-40^{\circ}51'29''.9$ & - & $ \approx 6.2$* & - & 1 & - & [27], [28], [29]\\
IGR J17062-6143 & CPT & $17^{h}06^{m}16^{s}.3$ & $-61^{\circ}42'41''$ & - & $ \approx 7.3$* & $0.633$ & 3 & - & [30], [31]\\
4U 1708-23 & - & $17^{h}08^{m}23^{s}.0$ & $-22^{\circ}48'12''$ & - &$ \approx 10$ & - & 1 & - & [32]\\
4U 1705-44 & AR & $17^{h}08^{m}54^{s}.47$ & $-45^{\circ}06'07''.4$ & - & $6.6 \pm 1.1$ & - & - & 1 & [33], [34], [35]\\
SAX J1712.6-3739 & CT & $17^{h}12^{m}37^{s}.1$ & $-37^{\circ}38'40''$ & - & $4.8 \pm 1.0$ & - & 4 & - & [36], [37], [38]\\
RX J1718-4029 & C\textbf{\textsuperscript{\Cross}} & $17^{h}18^{m}24^{s}.1$ & $-40^{\circ}29'30''$ & - & $6.1 \pm 0.4$ & - & 1 & - & [39], [40]\\
3A1715-321 & - & $17^{h}18^{m}47^{s}.40$ & $-32^{\circ}10'40''.0$ & - & $\approx 5.45$ & - & 1 & - & [41], [42]\\
IGR J17254-3257 & CT\textbf{\textsuperscript{\Cross}} & $17^{h}25^{m}25^{s}.5$ & $-32^{\circ}57'17''$ & - & $ \approx 14.5$* & - & 1 & - & [43], [44]\\
4U 1722-30 & ACG\textbf{\textsuperscript{\Cross}} & $17^{h}27^{m}32^{s}.9$ & $-30^{\circ}48'08''$ & $0.0 - 80.0$ & $7.40 \pm 0.5$ & - & 1 & - & [45], [46], [47]\\
KS 1731-260 & OT & $17^{h}34^{m}13^{s}.46$ & $-26^{\circ}05'18''.6$ & - & $4.6 \pm 0.7$ & - & - & 1 & [48], [49], [50]\\
Swift J1734.5-3027 & T & $17^{h}34^{m}24^{s}.2$ & $-30^{\circ}23'53''$ & - & $ \approx 7.2$* & - & 1 & - & [51], [52]\\
1RXH J173523.7-35401  & - & $17^{h}35^{m}23^{s}.0$ & $-35^{\circ}40'13''$ & - & $ \approx 9.5$* & - & 1 & - & [53]\\
SLX 1735-269 & C & $17^{h}38^{m}17^{s}.12$ & $-26^{\circ}59'38''.6$ & - & $5.8 \pm 0.9$ & $1.5$ & 4 & - & [54], [55], [56]\\
4U 1735-44  & AR & $17^{h}38^{m}58^{s}.3$ & $-44^{\circ}27'00''$ & $27.0 - 80.0$ & $7.2 \pm 1.6$  & $4.65$& - & 1 & [57], [58], [25]\\
SLX 1737-282 & C\textbf{\textsuperscript{\Cross}} & $17^{h}40^{m}42^{s}.83$ & $-28^{\circ}18'08''.4$ & - & $5.1 \pm 0.4$ & - & 3 & - & [59], [60]\\
XMM J174457–2850.3 & T & $17^{h}44^{m}57^{s}.3$ & $-28^{\circ}50'20''$& - & $6.5$* & - & 1 & - & [61], [62]\\
SAX J1747.0-2853 & T & $17^{h}47^{m}02^{s}.60$ & $-28^{\circ}52'58''.9$ & - & $4.5 \pm 0.9$ & - & 1 & 1 & [63], [64], [65]\\
SLX 1744-299 & CT\textbf{\textsuperscript{\Cross}} & $17^{h}47^{m}25^{s}.89$ & $-30^{\circ}00'01''.6$ & - & $ \approx 8.5$* & - & 3 & - & [66], [67]\\
GX 3+1 & A & $17^{h}47^{m}56^{s}.096$ & $-26^{\circ}33'49''.35$ & - & $4.4 \pm 0.5$ & - & 1 & 1 & [68], [69], [70]\\
EXO 1745-248  & DGT & $17^{h}48^{m}05^{s}.23$ & $-24^{\circ}46'47''.7$ & $35.0 - 39.0$ & $6.90 \pm 0.5$ & - & - & 1 & [71], [4], [72]\\
GRS 1747-312  & DEGT & $17^{h}50^{m}46^{s}.86$ & $-31^{\circ}16'28''.9$ & $74.0 - 90.0$ & $6.7 \pm 0.5$ & $12.4$ & 1 & - & [73], [74], [4]\\
AX J1754.2-2754 & AT & $17^{h}54^{m}14^{s}.50$ & $-27^{\circ}54'35''.6$ & - & $ \approx 7.1$* & - & 1 & - & [75], [76], [77]\\
SAX J1806.5-2215 & T & $18^{h}06^{m}32^{s}.168$ & $-22^{\circ}14'17''.32$ & - & $<4.8$ & - & 1 & - & [78], [79]\\
XTE J1810-189 & T & $18^{h}10^{m}20^{s}.86$ & $-19^{\circ}04'11''.2$ & - & $5.7 \pm 0.1$ & - & 1 & - & [80], [81], [82]\\
GX 17+2 & RZ & $18^{h}16^{m}01^{s}.39$ & $-14^{\circ}02'10''.6$ & - & $8.5 \pm 1.2$ & - & 14 & 4 & [83], [84], [85]\\
4U 1820-30  & ACGR & $18^{h}23^{m}40^{s}.5029$ & $-30^{\circ}21'40''.088$ & $34.0 - 81.0$ & $7.60 \pm 0.40$ & $0.19$ & - & 2 & [86], [87], [88], [4]\\
SAX J1828.5-1037 & T & $18^{h}28^{m}34^{s}.0$ & $-10^{\circ}36'59''$ & - & $<5.8$ & - & - & 1 & [89], [90]\\
GS 1826-238  & A & $18^{h}29^{m}28^{s}.2$ & $-23^{\circ}47'49''$ & $27.0 - 31.0$ & $6.7 \pm 0.2$ & $2.09$ & - & 1 & [91], [92], [93], [94]\\
Ser X-1 & AR & $18^{h}39^{m}57^{s}.55$ & $+05^{\circ}02'09''.5$ & $0 - 10.0$ & $7.8 \pm 1.4$ & - & - & 4 & [24], [95], [96]\\
4U 1850-086  & ACG & $18^{h}53^{m}04^{s}.88$ & $-08^{\circ}42'20''.0$ & $0 - 80.0$ & $ \approx 6.9$* & $0.343$ & 2 & - & [97], [98], [99], [4]\\
Aql X-1  & ADIORT & $19^{h}11^{m}16^{s}.047$ & $+00^{\circ}10'08''$ & $36.0 - 47.0$ & $4.2 \pm 0.3$ & $18.9$ & - & 2 & [100], [101], [102], [8]\\
M15 X-2 & C\textbf{\textsuperscript{\Cross}}G & $21^{h}29^{m}58^{s}.88$ & $+12^{\circ}10'02''.67$ & - & $6.6 \pm 0.5$ & - & 2 & - & [103], [104], [105], [4]\\
\hline
\textbf{Total: 40}& & & & & & & \textbf{56} & \textbf{28} \\

\hline
\end{tabular}
\\

{$^{a}$Source types; A = atoll, C = ultra-compact X-ray binary, D = "dipper", E = eclipsing, G = globular cluster association, I = intermittent pulsar, O = burst oscillation, P = accreting millisecond X-ray pulsar, R = radio-load X-ray binary, T = transient, Z = Z-source.\\
* These distances are adopted from other literature that reported the first PRE burst. The MINBAR-distances quoted are the inferred distances for a neutron star photosphere with hydrogen content of $X = 0$. \textbf{** Intermediate-duration burst}. \textbf{*** Superburst}  
\textbf{\textsuperscript{\Cross}} These sources are candidate ultra-compact binaries with indirect identifications (i.e. no orbital period measurements).
}\\
\textbf{Ref.}: [1] - \cite{Gal05}, [2] - \cite{kuin15}, [3] - \cite{forjo76}, [4] - \cite{Kuul03}, [5] - \cite{fiocchi11}, [6] - \cite{Jonker01}, [7] - \cite{Juett}, [8] - \cite{Cutri03}, [9] - \cite{zhongwang11}, [10] - \cite{Piro97}, [11] - \cite{Bas06}, [12] - \cite{intZ08}, [13] - \cite{Mason80}, [14] - \cite{Cou86}, [15] - \cite{Boi03}, [16] - \cite{Iar07}, [17] - \cite{Belian}, [18] - \cite{Cani80}, [19] - \cite{Chevalier89}, [20] - \cite{Belian76}, [21] - \cite{Wachter02}, [22] - \cite{Keek08}, [23] - \cite{Swank76a}, [24] - \cite{Asai00}, [25] - \cite{Casa06}, [26] - \cite{Russell12}, [27] - \cite{Homan07}, [28] - \cite{Kaplan08}, [29] - \cite{falan09}, [30] - \cite{degenaar12}, [31] - \cite{Stroh18}, [32] - \cite{Hoffman78}, [33] - \cite{sztajno85}, [34] - \cite{disalvo05}, [35] - \cite{Piraino07}, [36] - \cite{Cocchi99}, [37] - \cite{cummings14}, [38] - \cite{fiocchi08}, [39] - \cite{Kaptein}, [40] - \cite{intZ05a}, [41] - \cite{tawaraa}, [42] - \cite{Kuul10}, [43] - \cite{Brandt06}, [44] - \cite{Chenevez07}, [45] - \cite{Swank}, [46] - \cite{Grindlay80}, [47] - \cite{Kuul03}, [48] - \cite{Sunyaev89}, [49] - \cite{Cack06}, [50] - \cite{Zur10}, [51] - \cite{Ken13}, [52] - \cite{bozzo15}, [53] - \cite{degenaar10}, [54] - \cite{Baz97}, [55] - \cite{dav97}, [56] - \cite{Wilson03}, [57] - \cite{Lew77}, [58] - \cite{Aug98}, [59] - \cite{tomsick07}, [60] - \cite{intZ02}, [61] - \cite{Sak05}, [62] - \cite{degenaar12b}, [63] - \cite{intZ98}, [64] - \cite{Wijnands02}, [65] - \cite{Werner}, [66] - \cite{Pav}, [67] - \cite{ZoRe11}, [68] - \cite{Mak83}, [69] - \cite{oost01}, [70] - \cite{vanberg14}, [71] - \cite{Mak81}, [72] - \cite{Trem15}, [73] - \cite{intZ00}, [74] - \cite{intZ03b}, [75] - \cite{chelovekov07a,chelovekov07b}, [76] - \cite{Bas08}, [77] - \cite{Armas13}, [78] - \cite{chakrabarty10}, [79] - \cite{Kaur17}, [80] - \cite{chakrabarty08}, [81] - \cite{Mark1443}, [82] - \cite{Krimm08}, [83] - \cite{Oda81}, [84] - \cite{farinelli07}, [85] - \cite{liu07}, [86] - \cite{Grindlay75}, [87] - \cite{Anderson97}, [88] - \cite{dtrigo17}, [89] - \cite{Cor02a}, [90] - \cite{Hands04}, [91] - \cite{Barret95}, [92] - \cite{u97}, [93] - \cite{Homer98}, [94] - \cite{intZ99a}, [95] - \cite{Swank76b}, [96] - \cite{Migliari04}, [97] - \cite{Hoffman80}, [98] - \cite{lehto90}, [99] - \cite{Homer96}, [100] - \cite{Lew76}, [101] - \cite{Welsh00}, [102] - \cite{Cam03}, [103] - \cite{vanParadijs90}, [104] - \cite{White01}, [105] - \cite{dieball05}
\label{tab:overview}
\end{table}
\twocolumn


\begin{equation} \label{eq5}
    F(t) = F_{\rm 0}\left(\frac{t - t_{\rm s}}{t_{\rm 0} - t_{s}}\right)^{-\alpha}
\end{equation}
where t is time, $F_{\rm 0}$ is the highest-measured bolometric flux at $t = t_{\rm 0}$, t$_{\rm s}$ is the time when cooling starts and $\alpha$ is the power law decay index.  
We then obtain \textit{E$_{\rm b}$} by integrating equation \ref{eq5} from the observed start time, $t_{\rm 0}$, of the burst 
to the (extrapolated) time when the bolometric flux returns to the pre-burst persistent flux.
However, if the onset of a burst is not observed, \textit{E$_{\rm b}$} remains a lower limit to the actual burst fluence. The estimated uncertainty on this parameter is obatained by evaluating the uncertainties on the fitted parameters of Eq. \ref{eq5}.

The burst timescale $\tau$ (in seconds) is the ratio of the bolometric fluence to peak bolometric flux, $F_{\rm peak}^{\rm bol}$ (eq. \ref{eq1}):
\begin{equation} \label{eq3}
\begin{split}
 \tau &= \frac{E_{\rm b}}{F_{\rm peak}^{\rm bol}} 
 \end{split}
\end{equation}
An estimate of the ignition depth based on the burst energy, y$_{\rm b}$ (in \gcm2), is given by:
\begin{equation} \label{eq4}
    y_{\rm b} = \frac{E_{\rm tot} \times (1 + z)}{4\pi R^{2}Q_{\rm nuc}}
\end{equation}
Here, $z$ is the gravitational redshift ($z = 0.24$ for a neutron star of mass $M = 1.4 M_{\odot}$ and radius $R = 12$~km), and $Q_{\rm nuc}$ is the energy release per unit weight ($0.2~\times~10^{18}$~erg~g$^{-1}$ for C, and $1.31~\times~10^{18}$~erg~g$^{-1}$ for He (\cite{Goodwin19})). 
The values for $z$, $M$, $R$, and $Q_{\rm nuc}$ are canonical and contributes with a systematic uncertainty of $\approx 3 \%$ to $y_{\rm b}$.

The $\gamma$ parameter is the ratio of the persistent bolometric flux (obtained through spectroscopy - see  \ref{subsubsec:persistent_spectral}) to the Eddington flux for the same source. The highest observed flux is used when no PRE bursts have been observed from a source. The uncertainty on this parameter is estimated by evaluating the uncertainties on the persistent bolometric flux and the Eddington flux.  

\begin{equation} \label{eq6}
    \gamma = 1.7\frac{F_{\rm pers}}{F_{\rm Edd}} \frac{\xi_{\rm p}}{\xi_{\rm b}}
\end{equation}
The factor $1.7 = 1+ X$ comes from the adopted composition of the Eddington-limiting atmosphere with a hydrogen fraction of $X = 0.7$. 
The ratio $\xi_{\rm p}/\xi_{\rm b}$ is the anisotropy correction ratio of the persistent flux to the burst peak flux (see eq. \ref{eq2}), that is $0.9$ for the non-dippers and $4.43$ for the dippers. 
If a source has not shown a verifiable PRE burst, we use the brightest observed burst from that particular source and give an upper limit of the $\gamma$-value.
Both the $1.7$ factor and the ratio $\xi_{\rm p}/\xi_{\rm b}$ are adopted from (\cite{Gal20}).

The last parameter is t$_{\rm PRE}$, defined, from the time-resolved spectral analysis, as the time from when the bolometric flux has peaked, to when the black-body temperature has peaked (i.e. the "touchdown point" \citep[e.g., ][]{Lew84, tawaraa, Gal_keek17}). For the \textit{Swift} bursts, we define t$_{\rm PRE}$ from when the rise begins in the BAT light curve to when the count rate peaks, which we assume to coincide with the end of the PRE phase (\cite{Keek17}). The uncertainty on this parameter is estimated based on the binning time of the time-resolved spectral analysis at the peak of the bursts.

\subsection{Cross-calibration with previous results}
\label{subsec:crosscalib}
The results presented in this study are all cross-calibrated with those previously published. 
In particular, results available in MINBAR are confirmed (within error margins) with some exceptions. Our analysis differs from that of MINBAR for the $\gamma$ ratio, as we 
extrapolate a bolometric flux (between $0.1 - 100$ keV) using the \texttt{XSPEC} model \texttt{CFLUX} from the pre-burst spectrum, while in MINBAR the persistent flux is measured in the $3 - 25$ keV band and corrected with source-specific bolometric corrections. 
This leads to marginally different $\gamma$-values (max. difference of $5$\%).
The basic parameters describing JEM-X bursts included in MINBAR are determined from the count-rate light curves, while these are determined from the time-resolved spectral analysis in our study.       

$F_{\rm peak}$ and $\gamma$ of the PCA bursts from GX~17$+$2 in Table \ref{tab:full}, are different from those previously published in (\cite{kuul02b}). We obtain the peak bolometric black body flux, $F_{\rm peak}$, from the 1-second peak spectrum, while the previous study uses 0.25~s spectra to obtain the peak flux. For the one burst detected in Standard 2 data mode, we apply equation \ref{eq1} to obtain our peak bolometric black body flux. Consequently, this gives us lower values for the peak flux, and thereby higher $\gamma$ than reported by (\cite{kuul02b}). In general, we have calculated $\gamma$ using the peak flux of a PRE burst either from our catalogue or from a PRE-flagged burst in MINBAR (\cite{Gal20}). For the GS~1826$-$238 superburst, we used the peak flux from the only PRE burst observed in 2014 by \textit{NuSTAR} (\cite{Chenevez16}). 

For the bursts not entirely covered by the observations, estimates of $E_{\rm tot}$ (for the calculation of $y_{\rm b}$) have been obtained by interpolating (and extrapolating) the fragmented bolometric light curves to a power law (see equation \ref{eq5}). In earlier publications \citep[e.g.][]{Cor02, kuul02a, Keek08}  the $E_{\rm tot}$ was obtained by fitting with an exponential decay. This results in different values of $E_{\rm tot}$ in our study than the published ones. Furthermore, we use $Q_{\rm nuc} = 1.31~\times~10^{18}$~erg~g$^{-1}$ (\cite{Goodwin19}) instead of $Q_{\rm nuc} = 1.6~\times~10^{18}$~erg~g$^{-1}$ for intermediate-duration bursts and assume a carbon fraction of $X_C = 0.2$ for superbursts. Futhermore, we apply the distances adopted from MINBAR \citep[][]{Gal20}, which changes $y_{\rm b}$) up to by a factor of $10$ from those previously published \citep[][]{C06, intZ17b}. 

The differences in parameter values with the previous results do not give rise to a different physical explanation for the bursts in our catalogue.

\section{The long-burst catalogue}
\label{sec:catalogue}
\subsection{Source characteristics}
\label{subsec:sources}
Table \ref{tab:overview} presents the general characteristics of the 40 sources with at least one confirmed long burst. Seventeen are transients, thirteen are atoll sources, one Z-source, fourteen UCXBs \citep[][]{Rappaport, nelemans, intZ07}, two Accreting millisecond X-ray pulsars (AMXP) (\cite{disalvo20}), seven have a globular cluster association, four have shown burst oscillations, eight have the designation radio-loud X-ray binary with radio to X-ray luminosity ratio $\geq 0.7$ (\cite{Gallo12}), and four are so-called dippers (\cite{White95}). The columns R.A. and Dec. are adopted from MINBAR (\cite{Gal20}). The MINBAR positions for the listed sources are the most precise known. The distance, d, to the sources (column 6) are also adopted from MINBAR, apart from the entries in the column marked with $^{*}$, which are our best estimates of the distances using PRE bursts as standard candles. The MINBAR distances listed in Table \ref{tab:overview} are inferred for a neutron star photosphere with a hydrogen mass fraction of $X = 0.0$, a neutron star mass of 1.4 $M_{\odot}$ and an observed (gravitationally redshifted) Eddington limit of $3 \times 10^{38}$ \ergps for a neutron star radius of $12$ km. Inclination angles are listed in column 5 for 14 of the sources.

In the following, we present in Table \ref{tab:full} a catalogue consisting of 70 long bursts previously published (through refereed papers) or reported (mostly through ATels), but re-analyzed systematically for this study.
In addition, we compile a literature study of 14 long bursts observed with an earlier generation of X-ray telescopes, but where the data is not available for analysis. The main properties of these bursts are retrieved from the original works and presented in Table \ref{tab:lit}.

\subsection{Re-analyzed X-ray bursts}
\label{subsec:re-ana}
\subsubsection{Table format}
\label{subsubsec:tab_form}
The properties listed in Table \ref{tab:full} are obtained by re-analyzing published data following the analysis procedures described in Section \ref{sec:data_analysis}. Here we describe the entries of Table \ref{tab:full} and how they are obtained. 

The $1^{st}$ column, "Source", denotes the burst source ordered according to R.A. (see Table \ref{tab:overview}). The \textbf{(S)}, next to some source names, designates superbursts. A total of 28 superbursts are included.

The $2^{nd}$ and $3^{rd}$ columns list the Modified Julian Date (MJD) of the bursts. 
The $4^{th}$ column lists the instrument(s) by which a specific burst is observed. 
The $5^{th}$ column indicates if the onset of a specific burst has been observed. All the satellite-specific date formats are first converted to MJD and then to Gregorian Date. 

The $6^{th}$ column, "$F_{\rm peak}^{\rm bol}$", lists the peak bolometric fluxes. Bursts observed with \textit{INTEGRAL}, \textit{Swift}, \textit{RXTE}/PCA, and \textit{NuSTAR} are all calculated using Equation \ref{eq1}, while the values for bursts observed with \textit{RXTE}/ASM and \textit{MAXI} are the maximum flux value observed, which is time-averaged over \textit{RXTE}/ASM dwell or a \textit{MAXI} scan. For sources with available inclination ranges (see Table \ref{tab:overview}), the peak flux is multiplied by the anisotropy factor, $\xi_{\rm b}$ (see Equation \ref{eq2}.)

The $7^{th}$ column lists the burst timescale, $\tau$, which is calculated using Equation \ref{eq3}. The energy fluence $E_{\rm b}$ is either directly measured from bursts fully covered by observation or integrated over the data gaps using Equation \ref{eq5}.

The $8^{th}$ column lists the $\gamma$ parameter, calculated with Equation \ref{eq6}. The pre-burst persistent flux is modelled using the procedure described in \ref{subsubsec:persistent_spectral} for bursts with data available less than 24 hours before or after the burst. The MINBAR database (\cite{Gal20}) was used in combination with our measurements to obtain the source Eddington fluxes.

The $9^{th}$ column lists the duration of the PRE phase, $t_{\rm PRE}$, as defined in \ref{subsubsec:burst_param}. The error on $t_{\rm PRE}$ is limited by the time resolution of our time-resolved spectral analysis at the touchdown point. For the \textit{Swift} bursts where the BAT instrument observes the PRE phase, the error is limited by the time resolution of the BAT light curve.

The $10^{th}$ column lists the observed fluence, $E_{\rm obs}$, obtained by integrating over the time intervals of the time-resolved spectral analysis. For bursts that are fully covered by observation, $E_{\rm obs}$ is the same as the total energy fluence. For bursts with data gaps, $E_{\rm obs}$ is a lower limit to the total energy fluence.

The $11^{th}$ column lists the total irradiated energy, $E_{\rm tot}$, at a distance d from the burst source (see Table \ref{tab:overview}). For bursts that are fully covered by observation, $E_{\rm tot}$ is $E_{\rm obs}$ integrated over a sphere with radius d. For bursts not fully covered by observation, we interpolate over data gaps, using Equation \ref{eq5}.

The $12^{th}$ column lists the ignition depths, $y_{\rm b}$, calculated with Equation \ref{eq4}. For intermediate-duration bursts, we assume pure helium burning with a total energy released per mass unit of $Q_{\rm nuc} = 1.31~\times~10^{18}$~erg~s$^{-1}$ (\cite{Goodwin19}). For superbursts, we assume a carbon fraction of $0.2$, resulting in $Q_{\rm nuc} = 0.2~\times~10^{18}$~erg~s$^{-1}$.

The $13^{th}$ and $14^{th}$ columns list the MINBAR IDs of bursts included in MINBAR (\cite{Gal20}) and references to previous works, respectively.

\subsubsection{General statistics}
\label{subsubsec:general}
The 70 long bursts in Table \ref{tab:full} consists of 42 intermediate-duration bursts and 28 superbursts. Here we mention general statistics of some of the entries described in \ref{subsubsec:tab_form}. 

The shortest burst timescale, $\tau = 76 \pm 7$~s, is measured for the 2013 intermediate-duration burst from SLX~1744$-$299, while the longest burst timescale, $\tau = 41416 \pm 3796$~s, is measured for the first superburst from 4U 1636$-$536. 

Twenty-two intermediate-duration bursts are observed from UCXBs, nine are observed from transients, while two bursts are observed from sources that are yet to be characterized. The remaining nine intermediate-duration bursts are observed from the Z-source GX~17$+$2. Fifteen superbursts are observed from atoll-sources with no UCXB association, four superbursts are observed from atoll-sources with UCXB association, three superbursts are observed from transients, and two superbursts are observed from sources yet to be characterized. The remaining four superbursts are observed from GX~17$+$2.    

The pre-burst and post-burst flux information are only available for 49 out of the 70 bursts. We, therefore, calculate $\gamma$ for these 49 bursts (26 intermediate-duration bursts and 23 superbursts). The range for $\gamma$ spans from $0.15 \%$ of the Eddington flux up to super-Eddington levels. 

As the triple-$\alpha$ reaction rates are fast, the fuel burns quickly and the luminosity reaches (and surpasses) the local Eddington limit for pure He bursts \citep[e.g.][]{bildsten95, intZ07}. We can therefore assume that all intermediate-duration bursts reach the Eddington luminosity, since they occur due to the ignition of an unusually thick He layer. Observationally, we have identified 37 intermediate-duration bursts that have a clear PRE phase, based on the anti-correlation between the black body temperature, $kT_{\rm BB}$, and the black-body radius. For the \textit{Swift} bursts we identify the end of the PRE phase as the "touchdown point" which coincides with the peak in the BAT count rate (see Section \ref{subsubsec:burst_em}). Of the remaining three bursts, two are observed with \textit{MAXI}, where the source is in the FOV for 50~s (average duration of a scan) every 90~min and thus preventing us from observing the PRE phase. The last burst is the two-phased burst from GX~3$+$1, which is discussed in Section \ref{subsec:peculiar}. The shortest PRE phase we observe is t$_{\rm PRE} = 10 \pm 2$~s, while the longest is t$_{\rm PRE} = 1000 \pm 60$~s of the 37 bursts where this is measured. Even though some precursors of superbursts may reach the Eddington luminosity, we do not measure the parameter t$_{\rm PRE}$ as the PRE phase is too short and the time resolution of the instruments in question is too low.

The total radiated energy for intermediate-duration bursts spans from $E_{\rm tot} = (1.24 \pm 0.05) \times 10^{40}$ erg (\textit{RXTE}/PCA burst from GRS~1747$-$312 (\cite{intZ03a})) to $E_{\rm tot} = (49 \pm 5) \times 10^{40}$~erg (the 2014 burst from SAX~J1712.6$-$3739 \citep[][]{cummings14, intZ19, lin20}). For superbursts, $E_{\rm tot}$ spans from  $(4.4 \pm 0.2) \times 10^{40}$~erg (ASM superburst from 4U~0614$+$091) up to $(240 \pm 40) \times 10^{40}$~erg (first \textit{INTEGRAL}/JEM-X from SAX~J1747.0$-$2853). Both intermediate-duration burst from SAX~J1712.6$-$3739 and the superburst from 4U~0614$+$091 are discussed further in Section \ref{subsec:peculiar}.

The ignition depth, $y_{\rm b}$, ranges from $(0.89 \pm 0.01) \times 10^{9}$\gcm2 up to $(39 \pm 4) \times 10^{9}$\gcm2 for intermediate-duration bursts, and from $(78 \pm 21) \times 10^{9}$\gcm2 up to $(1251 \pm 209) \times 10^{9}$\gcm2 for superbursts in our catalogue, assuming a carbon fraction $X_{\rm C} = 0.2$.   

We measure the observationally-inferred recurrence times for bursts from 13 out of the 40 sources in our catalogue. The shortest measured recurrence time for intermediate-duration bursts is $172$ days from SLX~1744$-$299, while the longest is $2073.12$ days for the two bursts from 2S~0918$-$549. The shortest measured recurrence time for superbursts is $363.8$ days for the four superbursts from GX~17$+$2, and the longest is $1761.5$ days for the two superbursts from 4U~0614$+$091. A cluster of five intermediate-duration bursts was observed from GX~17$+$2 in October 1999, with an inferred recurrence time of $1.4$ days.

\subsection{Main trends}
\label{subsec:main_trends}

\begin{figure}
\centering
\subfloat{\includegraphics[width = 3.2in]{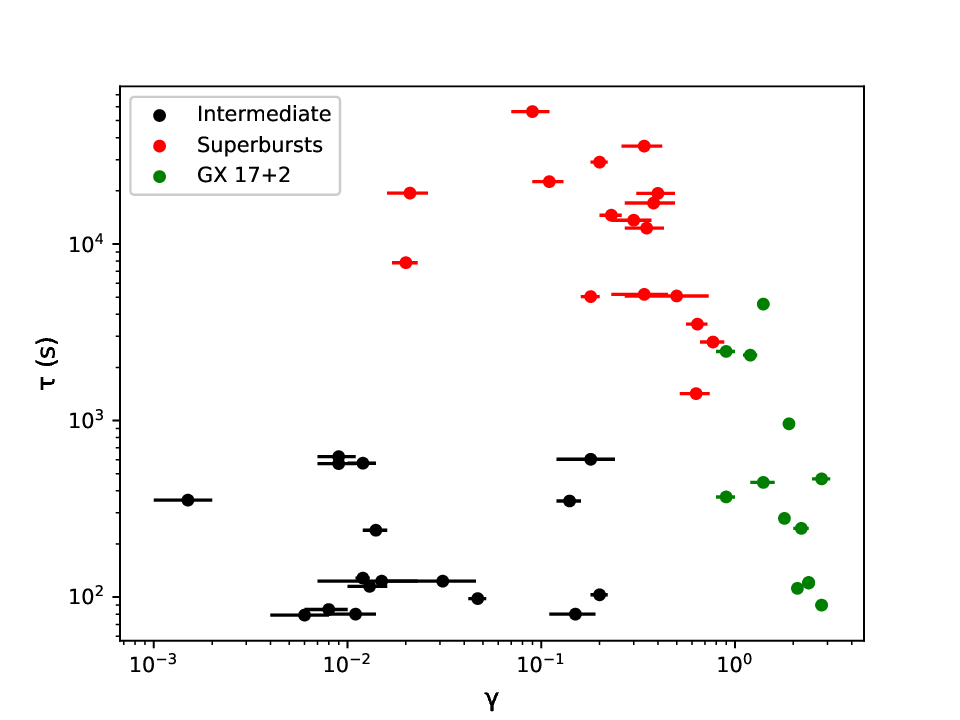}}\\  
\caption{The persistent emission $\gamma$ against the burst timescale $\tau$(s) for 49 bursts from Table \ref{tab:full}.}
\label{Fig:gamma}
\end{figure}

Here we discuss the main trends that appear from our catalogue.
In Figure \ref{Fig:gamma} we compare the persistent emission, $\gamma$, with the burst timescale, $\tau$, revealing the following trends. 
Superbursts have all $\tau \gtrsim 1000$~s, while intermediate-duration bursts have not.
These latter occur at $\gamma$-values that ranges from $0.006$ up to $0.29$ (excluding bursts from GX~17$+$2, which are discussed in Section \ref{subsec:peculiar}) with the majority around $\gamma = 0.01$, consistent with a low accretion rate from a Helium-rich companion \citep[e.g.;][]{intZ11, intZ12, degenaar18}. 
Superbursts occur at $\gamma$-values ranging from $0.02$ up to $0.77$ (excluding the four superbursts from GX~17$+$2), with a clear majority of the superbursts occurring at $\gamma \gtrsim 0.3$. 
There is thus no clear distinction between intermediate-duration bursts and superbursts regarding the relative persistent fluxes at the time they arise. However, there is a trend for superbursts to take place at accretion rates above 30\% of Eddington, where intermediate bursts are not present.
 
\begin{figure}
\centering
\subfloat{\includegraphics[width = 3.2in]{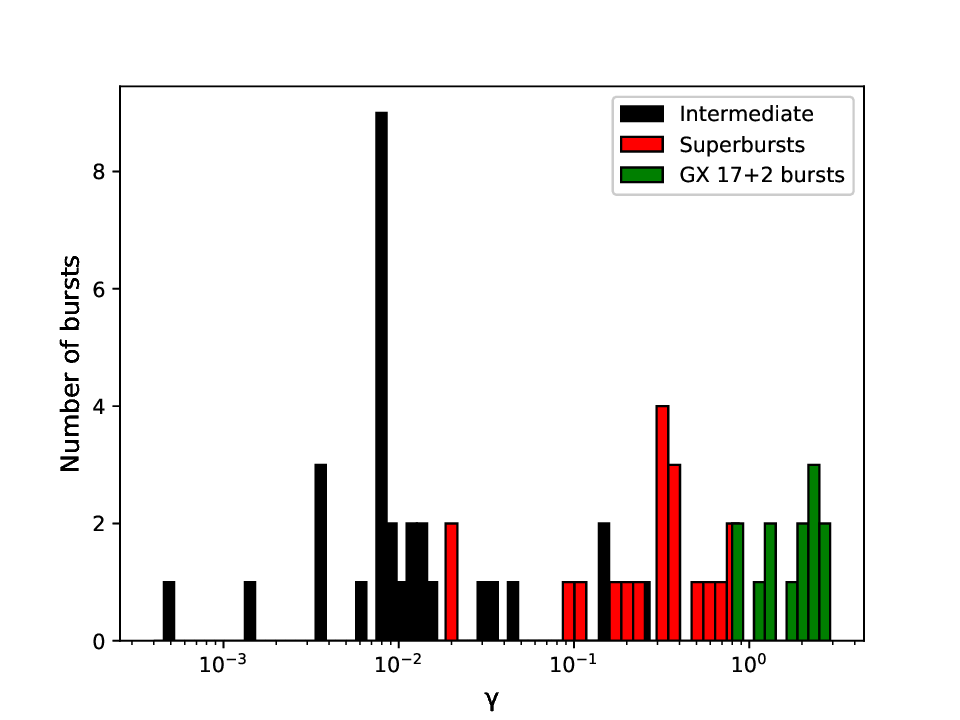}}\\  
\caption{Long burst distribution as a function of the persistent flux, $\gamma$, for 64 bursts including 15 additional bursts from slow accretors (see text).
}
\label{Fig:pers}
\end{figure}

The following six UCXBs and two transient sources IGR~J17962$-$6143, SAX~J1712$-$3739, XMM~J174457$-$2850.3, SLX~1744$-$299, SAX~J1806.5$-$2215, XTE~J1810$-$189, 4U~1850$-$086, and SLX~1735$-$269 are all slow accretors 
that do not show a significant persistent flux variations over long periods. We can thus infer the corresponding $\gamma$ values for 15 bursts from these sources. To further clarify the distribution of long bursts, we plot in Figure \ref{Fig:pers} a histogram of the persistent flux for 64 bursts (including the 49 bursts in Fig. \ref{Fig:gamma} and the 15 additional bursts).  
We see that the majority of the intermediate-duration bursts occur at $\gamma \approx 0.01 - 0.03$ with some significant outliers, while superbursts occur at $\gamma \gtrsim 0.2$ also with a couple of outliers. 
In comparison to long bursts, classical bursts occur at all values of $\gamma$ in the range of $10^{-3 - 4}$ \citep[e.g.;][]{Gal20}.

\begin{figure}
\centering
\subfloat{\includegraphics[width = 2.2in, angle=-90]{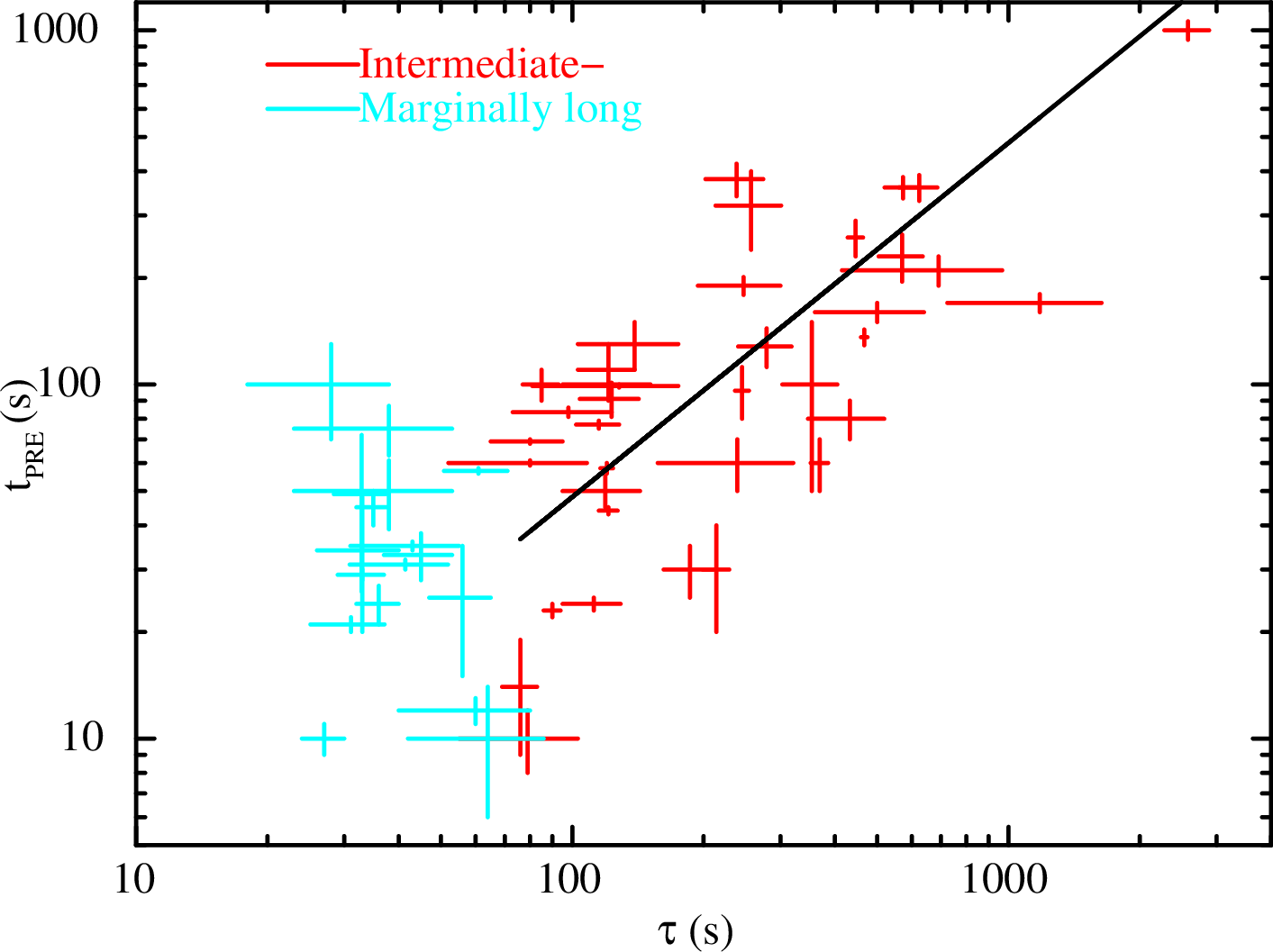}}\\  
\caption{Burst timescale $\tau$ against the duration of the PRE phase $t_{\rm PRE}$. The red data points represent the intermediate-duration bursts, while the cyan data points represent the marginally-long bursts discussed in Section \ref{subsec:marginal}. The black line represents the best linear fit to the intermediate-duration bursts with a slope value of $\alpha = 0.48 \pm 0.01$. }
\label{Fig:PRE}
\end{figure}

In Figure \ref{Fig:PRE} we plot the duration of the radius expansion phase $t_{\rm PRE}$ as a function of the burst timescale $\tau$. The red data points in Figure \ref{Fig:PRE} represent 37 intermediate-duration from Table \ref{tab:full} for which we have measured $t_{\rm PRE}$. The black line in Figure \ref{Fig:PRE} is the best-fit linear relation for the intermediate-duration bursts with a proportionality factor $a = 0.48 \pm 0.01$, indicating that on average, half of the effective burst duration, $\tau$, is attributed to the PRE phase. 
\citet{intZ10} previously found a proportionality factor $\approx 1$ between $\tau$ and $t_{\rm PRE}$ for a collection of bursts with a confirmed super-expansion phase. The discrepancy between this study and the proportionality factor obtained in Fig. \ref{Fig:PRE} may be due to the bursts included in \cite{intZ10} all are relatively short with $30 \leq \tau \leq 50$~s for the majority of them and the $\tau$ values in the previous study were obtained from exponential fits. We do not detect the presence of super-expansion phases in the bursts from our sample. The cyan data points represent $t_{\rm PRE}$ for marginally-long bursts discussed in Section \ref{subsec:marginal}. However, for three marginally-long bursts, two observed with \textit{BeppoSAX} (from SLX~1744-299 and 4U 0513-40) and one with \textit{Swift} (from IGR~J18245$-$2452), the burst timescale $\tau$ is smaller than the duration of the PRE phase $t_{\rm PRE}$. The discrepancy between $\tau$ and $t_{\rm PRE}$ for the \textit{Swift} burst is caused by the shortened tail due to missing data, while the same discrepancy for the \textit{BeppoSAX} bursts may be due to the low SNR of the instrument preventing us from observing the whole tail of the burst. 

\begin{figure}
\centering
\subfloat{\includegraphics[width = 3.4in]{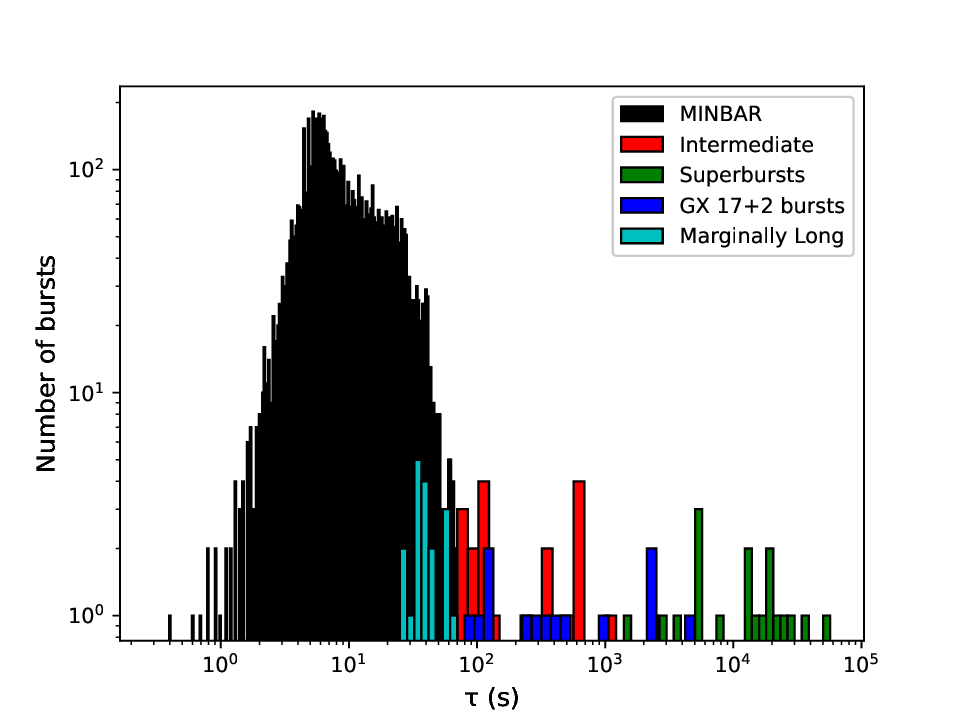}}\\  

\caption{Type-I X-ray burst distribution as a function of the burst timescale $\tau$. The distribution consist of 7013 bursts from MINBAR, 33 intermediate-duration bursts, 24 superbursts, 13 bursts from GX~17$+$2 and 18 marginally-long bursts.
} 
\label{Fig:hist.}
\end{figure}

In Figure \ref{Fig:hist.} we compare the distribution of burst timescales from the four groups 
(intermediate-duration bursts, superbursts, GX~17$+$2 bursts, and marginally-long bursts) discussed in this study, with the (short) MINBAR bursts. We have applied an upper limit of $\tau = 70$~s and filtered out the bursts with no burst-timescale information in MINBAR, reducing the sample from 7083 to 7013 bursts. From Figure \ref{Fig:hist.} we see a clear distinction between the burst timescales of intermediate-duration bursts and superbursts. The bursts from the Z-source GX~17$+$2 are depicted separately, as they show a divergent behaviour compared to other long bursts in our catalogue (see Section \ref{subsec:peculiar}). 

\begin{figure}
\centering
\subfloat{\includegraphics[width = 3.4in]{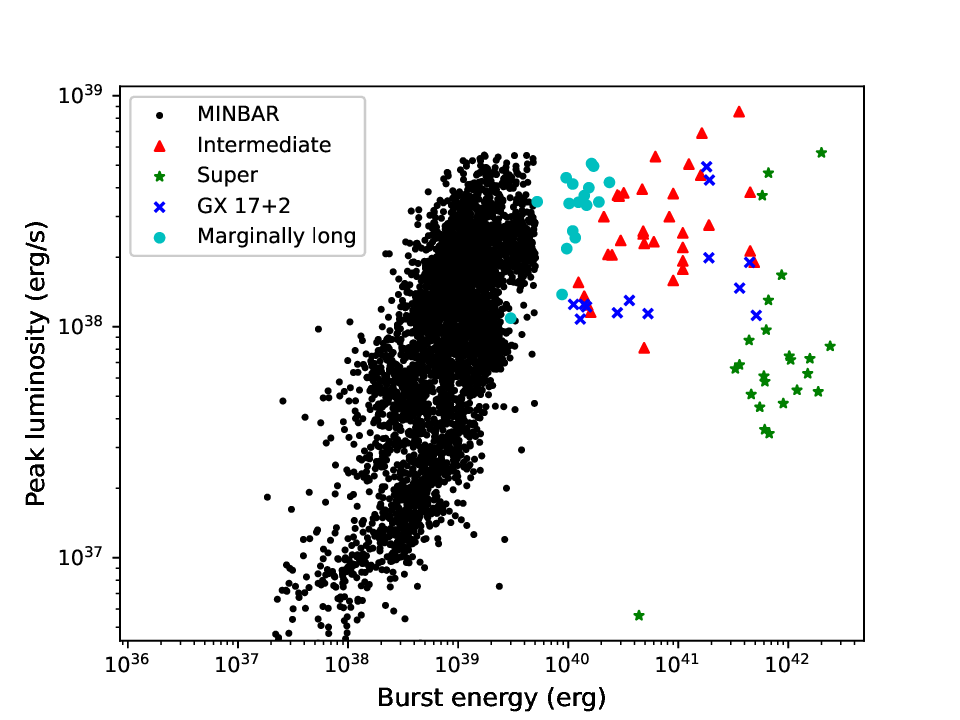}}\\  
\caption{Peak bolometric luminosity as a function of the total irradiated energy $E_{\rm tot}$ during a burst.}
\label{Fig:etot}
\end{figure}


\subsection{Bursts from the literature}
\label{subsec:literature}
Although observations of long bursts have especially been of high interest in the last couple of decades, the first observations of these rare events were made with the early generation of X-ray instruments. For the sake of completness, we give here an overview of 14 long bursts detected mainly between 1969-1990 (11 bursts) or later but whose data have become unusable (3 bursts). We have used measurements from the previously-published works and estimated parameters needed to make these literature bursts compatible with the rest of the present catalogue (see Table \ref{tab:lit} for burst parameters). These 14 bursts were observed by the following instruments: \textit{Hakucho} (0.1 -- 100 keV), \textit{SAS-3} (0.1 -- 60 keV), \textit{Einstein} (0.15 -- 20 keV), \textit{Ginga}/LAC (1.5 -- 37 keV), Vela 5B (3 -- 750 keV), \textit{GRANAT} (2 keV -- 1.3 MeV), \textit{EXOSAT} (0.04 -- 80 keV), JEM-X (3 -- 35 keV, in restricted imaging data format) and \textit{HETE}/FREGATE (6 -- 400 keV).

\subsubsection{GX 17+2}
\label{subsubsec:GX17+2}
\cite{tawarac} report on four bursts detected from GX~17+2 by \textit{Hakucho} with durations ranging from $3-15$ min. These authors measured the peak flux of the bursts from the count-rate light curves over an energy range of $1-22$~keV and the persistent flux of the source from two observations (not coinciding with the bursts) over periods of $40$ hr and $50$ hr.
We assume the events to be \textit{FRED} (fast rise exponential decay) bursts and use the measured peak fluxes to calculate the burst energy release and ignition depths by integrating over the exponential curve (see Table \ref{tab:lit}).
\cite{sztajno} report a $\approx 5$ min long burst from GX 17+2 observed on 20 August 1985 by \textit{EXOSAT}. 
We consider the highest flux derived from their time-resolved spectral analysis as a lower limit for the bolometric peak flux. From there, we estimate the burst energy release and ignition depth by integrating over an exponential curve (using $t_{\rm exp}$) and the distance given in Table \ref{tab:overview}. 

\subsubsection{M15 X$-$2} 
\label{subsubsec:M15}
M15 X-2 is an UCXB located in the globular cluster M15. On 20 October 1988, the Large Array Counter (LAC) on-board Ginga observed a $3$-min long burst from this source. It is noted in \cite{vanParadijs90} that the burst showed a clear radius expansion phase lasting more than $40$~s at the peak. The long PRE phase is the main reason for including this burst in our catalogue. 
The observed burst peak flux is not reported by \cite{vanParadijs90}, but based on their time-resolved spectroscopy results, these authors derive an Eddington peak flux about $3.6 \times 10^{-8}$\ergcs. 
\cite{vanParadijs90} also quote a source distance of $9.2$~kpc based on observations of RR Lyrae stars.
We note that another long burst from this source was detected by WFC on 23 November 2000 (see Table \ref{tab:full}) and is also included in MINBAR \citep{Gal20}.

\subsubsection{4U~1722$-$30}
\label{subsubsec:4U1722}
4U~1722-30 is an UCXB candidate (\cite{intZ07}) located in the globular cluster Terzan 2. On 5 March 1979, the \textit{High Resolution Imager (HRI)} and the \textit{Monitor Proportional Counter (MPC)} on-board the \textit{Einstein X-ray observatory} observed an intermediate duration burst, with a $20$ s long radius expansion phase. In the original work (\cite{Grindlay80}), the peak luminosity is measured to be $5 \times 10^{38}$\ergps, well beyond the Eddington luminosity.  We use this PRE phase to estimate the distance to the source to be d $\approx 7.8$~kpc for $L_{Edd} = 3 \times 10^{38}$ \ergps. The distance estimate using RR~Lyrae stars of the cluster is $6.6$ kpc, which is similar to the distance $6.9 \pm 0.2$~kpc we estimate using our time-resolved spectral analyse of another PRE burst from the same source (see Section \ref{subsec:peculiar}).

\subsubsection{SLX~1744$-$299}
\label{subsubsec:SLX1744}
SLX~1744-299 is a burster located near the Galactic Center. Since its discovery in 1987 (\cite{skinner87}), there have been eight relatively long bursts (three of them do not meet our long burst criteria) detected from this source (the most recent in March 2020 by \textit{INTEGRAL}/JEM-X). Due to the angular proximity of the source to another X-ray burster, SLX~1744-300 (they are separated by less than $2.8$~arcmin), it has been very difficult to obtain any good data during its persistent phase. With angular resolutions worse than $\geq 3$~arcmin, one can still distinguish the two sources when one of them exhibits a burst. 
The first burst from this source was on 9 October 1990 observed by the ART-P coded-mask X-ray telescope on-board the \textit{GRANAT} observatory. This burst is the only known case of a PRE burst from this source. In the original work, a distance of $8.5$~kpc was estimated (see the parameters in Table \ref{tab:lit}).

\subsubsection{4U~1708$-$23}
\label{subsubsec:4U1708}
An intermediate burst with a duration of $t \geq 310$~s was detected by \textit{SAS-3} on 7 February 1977 (\cite{Lew84}). The source of the burst has been uncertain since the observation, but in the original work the transient 4U 1708-23 was proposed as a good candidate. 

The event is a PRE burst with a "precursor", separated from the "main" event by $\approx 5$~s. In the original work the authors assume a distance of $10$~kpc and show that the luminosity reaches the Eddington limit (see Table \ref{tab:lit}).

\subsubsection{3A~1715$-$321}
\label{subsubsec:3A1715}
The weak persistent source 3A~1715-321 ($F_{pers} = 8 \times 10^{-10}$\ergcs) was discovered in 1976 (\cite{markert76}) and Type-I bursts were detected soon after. 
On 20 July 1982, the third-ever long burst with a precursor was detected by the \textit{Hakucho} satellite. 
In the original work (\cite{tawaraa}; \cite{tawarab}), the derived peak luminosity is $8.0 \times 10^{38}$\ergps, well above the theoretical Eddington luminosity. 
We have here discarded that and instead used the Eddington luminosity $L_{Edd} = 3.0 \times 10^{38}$\ergps (for pure He bursts) to estimate the upper limit of the distance to the source  to be $d \approx 5.45$~kpc.

\subsubsection{Cen~X$-$4}
\label{subsubsec:CenX4}
The transient LMXB Cen X$-$4 is an X-ray burster that has been in quiescence since 1979 (\cite{vaneijnden21}). The first burst detected from this source is also the brightest X-ray burst ever because of the source proximity \citep[$\le 1.2$~kpc;][]{Kuul09}. It occurred on 7 July 1969 and was observed by the \textit{Vela 5B} satellite at a peak flux of $1.4 \times 10^{-6}$\ergcs.

\subsubsection{4U~0614$+$091 and SLX~1735-269}
\label{subsubsec:4U0614}
We include here three more bursts for which it has not been possible to reduce the data and thus perform our own analyses, though these events occurred after 2000.

The French Gamma-Ray Telescope (\textit{FREGATE}) on-board the \textit{High Energy Transient Explorer satellite (HETE-2)} detected on 17 February 2002 an intermediate-duration burst from the UCXB candidate 4U~0614+091 with an e-folding decay time of $89 \pm 5$~s obtained from fitting the $7-40$~keV light curve (originally reported in \cite{Kuul10}). 
Another burst reported by \cite{suzuki05} was detected by \textit{FREGATE}/WXM on 20 June 2005 from SLX~1735-269. The burst was observed for 24~min before the spacecraft made a slew. It was initially characterized as a superburst, but we denote it as an intermediate-duration burst since any other observatory has not confirmed that. 

\cite{molkov05} report on an intermediate-long burst from SLX~1735-269 detected with \textit{INTEGRAL} on September 2003. Unfortunately, the JEM-X data format used at the time of this event is now outdated, preventing us from reanalysing this data.
\label{subsection:literatur}

\section{Discussion}
\label{sec:discussion}
\subsection{Identification of long bursts}
\label{subsec:dataselec}
Our study presents the largest sample of long-duration X-ray bursts assembled to date. The bursts included in our catalog have been previously reported (either through regular publications or through shorter communications like \textit{The Astronomer's Telegrams (ATels)}), and we do not search here for new long bursts in archival data.  

Based on our study, we find that the following three criteria must be fulfilled for inclusion in our catalogue either as intermediate-duration bursts or as superbursts:

\begin{itemize}
    \item The total energy release through radiation, $E_{\rm tot}$, is larger than $10^{40}$~ergs.
    \item The burst exhibits PRE for longer than $10$~s.
    \item The burst timescale $\tau$ is longer than $70$~s (see Section \ref{subsubsec:burst_param}).
\end{itemize}

\begin{figure}
\centering
\subfloat{\includegraphics[width = 2.3in, angle=-90]{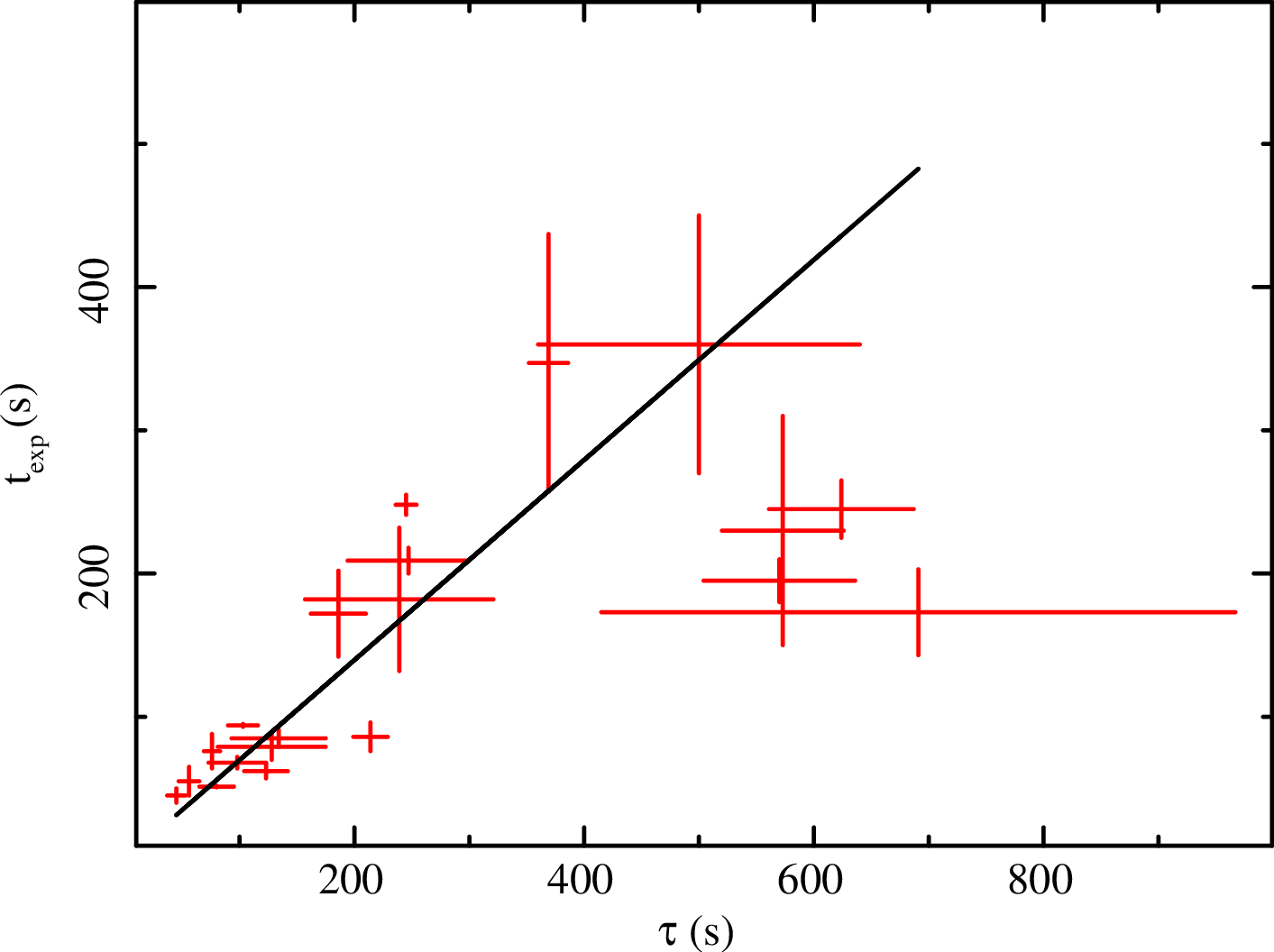}}\\  
\caption{ Burst timescale $\tau$ against the exponential decay time obtained from the count-rate light curves for 20 intermediate-duration bursts. The best-fit linear relation between the two parameters is over-plotted (black line). The slope value is $a = 0.70 \pm 0.01$.}
\label{Fig:tau_inter}
\end{figure}

The $\tau>70$~s criterion requires that bursts with long tails due to long rp‐process burning (\cite{Wallace81}) are excluded. This ensures that long bursts are only due to large ignition depths and corresponding long cooling times. Theoretically, the rp‐process can last up to 200~s after burst ignition (\cite{schatz01}). If we equalize this to the point in the tail of a burst where the burst flux drops below 1\% of the peak flux, the equivalent exponential decay time is 43~s. Applying a one-sided marginally error of $12\%$ at $95\%$ confidence level gives an exponential decay time of 50~s. In order to identify bursts with exponential decay times $t_{\rm exp} \approx 50$~s, we fit the count-rate light curve to an exponential and then draw an equivalency between exponential decay time and burst timescale. In Figure \ref{Fig:tau_inter}, we compare exponential decay times of 20 intermediate-duration bursts, obtained by fitting 3--25~keV band\footnote{assuming most of the radiation from the burst is in this band} count-rate light curves, to their burst timescales $\tau$. It appears that $t_{\rm exp} \approx 50$~s corresponds to a burst decay time scale of $\tau=70$~s, as we infer from a linear fit of exponential decay times versus burst decay time scales.


\vspace{0.5cm}

\subsection{Marginally long bursts}
\label{subsec:marginal}
As a consequence of our selection criteria (see Section \ref{subsec:dataselec}), there remain 18 bursts that are not part of the long-duration X-ray bursts listed in Table \ref{tab:full}. They only satisfy one or two of our three selection criteria.
These 18 bursts are gathered in Table~\ref{tab:short_bursts}, showing that they originate from 11 sources. Moreover, only the first two bursts (from 4U~0513-40) are included in MINBAR (IDs 2082 and 2094).
These bursts have a PRE phase $> 10$~s, making them unlikely to be powered by the rp-process, but they have burst timescales shorter than $70$~s. 
Comparing the burst timescales of all bursts (the long ones from the present study and the short ones from MINBAR) in Figure~\ref{Fig:hist.}, the 18 bursts overlap those of the longest bursts in the MINBAR sample, so we designate them as marginally-long bursts.

Generally, it has been the norm to clearly distinguish between classical bursts (He or mixed H/He ignition), intermediate-duration bursts (ignition of a thick layer of He), and superbursts (ignition of a deep layer of C). 
However, it is evident from observations that there are X-ray bursts that do not fit in any of these three conventional categories but can instead be placed between them, resulting in a continuous distribution of burst durations, at least for pure He burning. This continuous distribution of burst durations has previously been shown for the UCXB 4U~0614$+$091 through a systematic all-sky search for bursts from slow accretors using the \textit{FERMI}-GBM (\cite{linares12}). The continuous distribution of burst durations and energies from 4U~0614$+$09 was suggested to be a consequence of the C/O-dominated abundances of the donor star (\cite{Werner06}). Of the 11 sources listed in Table \ref{tab:short_bursts}, seven are UCXBs (one of them being 4U~0614$+$091), and four are transients. High neon-to-oxygen abundance ratios have been observed for 2S~0918$-$549 and 4U~1246$-$588, but these are attributed to ionization effects rather than the indicators of the donor compositions  \citep[][]{intZ05, intZ08}. 
In Fig. \ref{Fig:hist.}, there appears a group of bursts, including the marginally-long bursts, whose durations fall between the intermediate-duration bursts and the classical  bursts.

If we instead compare the distribution of the burst energies, we get a clearer picture. In Figure \ref{Fig:etot}, we plot the burst peak luminosity as a function of the total irradiated energy. The marginally-long bursts make a distinct population between the bursts from the MINBAR sample and the intermediate-duration bursts. The peak luminosities of the marginally-long bursts are also similar to those of intermediate-duration bursts. 
In Figure~\ref{Fig:etot}, it does also appear that four intermediate-duration bursts overlap the group of superbursts as regards their total energy. These bursts are those from IGR~J17062-6143 in 2015, SAX~J1712.6-3739 in 2014, SLX~1735-269 in 2012, and 4U~1850-086 in 2015. The two first ones also have $\tau$-values that overlap superbursts in Fig.~\ref{Fig:hist.}. Since the derived total energies are at most lower limits, it may be relevant to discuss if these four bursts should rather be considered as superbursts.
Unfortunately, the $\gamma$-value is only available for the burst from SLX~1735-269 (at $\gamma = 0.14$), which is more consistent with an intermediate-duration burst.

\subsection{Peculiar long bursts}
\label{subsec:peculiar}
The unusually long intermediate-duration burst observed in 2014 from SAX~J1712.6$-$3739 is not categorized as a short superburst because of the low accretion rate of the source, that is insufficient to produce the required quantities of C (\cite{C06}). However, the accretion rate of 4U~0614+091 was also very low $\gamma \sim 0.02$, when two superbursts were observed from it in 2005 and 2014 \citep[][]{Kuul10, Serino16}. In this case, one can argue that our current method of defining different bursts is inconsistent. Previous work did suggest ignition of an extremely thick He layer as an explanation for the 2005 superburst from 4U~0614$+$091 (\cite{Kuul10}), though, this explanation may also be applied to the 2014 superburst from the same source and the 2014 long burst from SAX~J1712.6$-$3739, suggesting that intermediate-duration bursts can in some cases rival the duration of superbursts. Assuming that the fuel for the two superbursts is pure He, then the ignition depths $y_{\rm b}$ are $2.43 \times 10^{9}$ and $37 \times 10^{9}$ \gcm2 for the 2005 and 2014 superbursts, respectively. These relatively shallow ignition depths indicate that the superbursts from 4U~0614$+$091 may be due to a thermonuclear runaway in an extremely thick helium layer, making them unusually long intermediate-duration bursts, which may also be the case for the 2014 burst from SAX~J1712$-$3739 \citep{intZ19}.  

The intermediate-duration burst from GX~3$+$1 observed by \textit{INTEGRAL}/JEM-X consists of two distinguishable phases, a classical He flash-like burst at the peak followed by a $\approx30$-minute cooling tail, which resembles the prolonged tail of the rp-process (\cite{Chenevez07}; \cite{alizai}) but continues well-above the theoretical limit for the rp-process timescale of $200$~s (\cite{schatz01}). Currently, the best explanation for this prolonged tail is an enhanced persistent emission component (\cite{Czerny87}).

As shown in Fig. \ref{Fig:gamma}, long bursts from the Z-source GX~17$+$2 occur at $\gamma$-values ranging from $0.9$ up to $2.8$. The burst emission is thought to be highly anisotropic, making the inferred persistent flux higher than the peak bolometric flux in some cases (see Table \ref{tab:full} and \cite{kuul02b}).
In Figure \ref{Fig:hist.}, the burst timescales for the 13 bursts from GX~17$+$2 extend over those of the other intermediate-duration bursts and superbursts, where the nine shortest bursts overlap intermediate duration bursts, while the longest bursts are located near the shortest superbursts from the other sources.
Following conventional thinking, the neutron star in GX~17+2 would not be cool enough to allow a significant amount of He to be accumulated. Nevertheless, six out of the nine intermediate bursts seen from this source show significant PRE phases, which indicates ignition of a thick layer of pure He. All the six bursts were observed by \textit{RXTE}/PCA. The remaining three bursts were observed by \textit{INTEGRAL}/JEM-X (two) and \textit{RXTE}/PCA in Standard 2 mode and did not show any PRE phase.


\onecolumn
\begin{landscape}
\scriptsize
\begin{xltabular}{\textwidth}{lccccccccccccc}
\caption{The primary parameters of 70 bursts included in this catalogue.} \\
\hline \multicolumn{1}{c}{\textbf{Source}} & \multicolumn{1}{c}{\textbf{MJD}} & \multicolumn{1}{c}{\textbf{Date}} & \multicolumn{1}{c}{\textbf{Instrument}} & \multicolumn{1}{c}{\textbf{Onset}} & \multicolumn{1}{c}{\textbf{$F_{\rm peak}^{\rm bol}$}$^{1}$} & \multicolumn{1}{c}{\textbf{$\tau$ (s)}} & \multicolumn{1}{c}{\textbf{$\gamma$}} & \multicolumn{1}{c}{\textbf{$t_{\rm PRE}$ (s)}} & \multicolumn{1}{c}{\textbf{$E_{\rm obs}$ $^{2}$}} & \multicolumn{1}{c}{\textbf{$E_{\rm tot}$ $^{3}$}} & \multicolumn{1}{c}{\textbf{$y_{\rm b}$}$^{4}$} & \multicolumn{1}{c}{\textbf{MINBAR-ID}} & \multicolumn{1}{c}{\textbf{Ref.}}\\ \hline 
\endfirsthead
\multicolumn{3}{c}%
{\tablename\ \thetable{} -- continued from previous page} \\
\hline \multicolumn{1}{c}{\textbf{Source}} & \multicolumn{1}{c}{\textbf{MJD}} & \multicolumn{1}{c}{\textbf{Date}}  & \multicolumn{1}{c}{\textbf{Instrument}} & \multicolumn{1}{c}{\textbf{Onset}} & \multicolumn{1}{c}{\textbf{$F_{\rm peak}^{\rm bol}$} $^{1}$ } & \multicolumn{1}{c}{\textbf{$\tau$ (s)}} & \multicolumn{1}{c}{\textbf{$\gamma$}} & \multicolumn{1}{c}{\textbf{$t_{\rm PRE}$ (s)}} & \multicolumn{1}{c}{\textbf{$E_{\rm obs}$}$^{2}$} & \multicolumn{1}{c}{\textbf{$E_{\rm tot}$ $^{3}$}} & \multicolumn{1}{c}{\textbf{$y_{\rm b}$}$^{4}$} & \multicolumn{1}{c}{\textbf{MINBAR-ID}} & \multicolumn{1}{c}{\textbf{Ref.}} \\ \hline  
\endhead \\
\hline \multicolumn{4}{r}{{continued on next page}} \\ \hline
\endfoot \\
\hline
\endlastfoot
IGR J00291-5934  & 57228.09&   2015-07-25  &   BAT/XRT  &   y  &  14.2 $\pm$ 3.6 &  79 $\pm$ 24 &  0.006 $\pm$ 0.002 &  10 $\pm$ 2 &  > 1.44 $\pm$ 0.24 &  2.11 $\pm$ 0.34 & 1.11 $\pm$ 0.18&  - &   [1]; [2] \\
4U 0614+091 \textbf{(S)}  & 53441.7&  2005-11-03  &  ASM  &   -  &  0.70  $\pm$ 0.08  &  7835 $\pm$ 290 &   0.020 $\pm$ 0.003  &  - &  > 2.4 $\pm$ 0.2 &  4.4 $\pm$ 0.2 & 15.12 $\pm$ 0.65&  - &   [3]; [4]; [5] \\
4U 0614+091 \textbf{(S)}  & 56964.63&  2014-11-03  &  MAXI  &   -  &  4.3 $\pm$ 0.5 &  19423 $\pm$ 2851 &  0.021 $\pm$ 0.005 &   -  &  > 11 $\pm$ 1 &  67 $\pm$ 6 & 229 $\pm$ 21&  - &  [4]; [5]\\
2S 0918-549   & 50357.88&   1996-10-01  &   WFC  &   y  &  13 $\pm$ 2 &  128 $\pm$ 47 &  0.012 $\pm$ 0.001 &  99 $\pm$ 1 &  17 $\pm$ 6 &  3 $\pm$ 1 & 1.58 $\pm$ 0.53& 1798&   [6] \\
2S 0918-549   & 54504.12&   2008-02-08  &   PCA  &   y  &  11.3 $\pm$ 0.9 &  115 $\pm$ 13 &  0.013 $\pm$ 0.003 &  77 $\pm$ 2 &  13 $\pm$ 1 &  2.3 $\pm$ 0.2 & 1.18 $\pm$ 0.13& 3663&   [7] \\
4U 1246-588  & 50461.29&   1997-01-13  &   WFC  &   y  &  14.6 $\pm$ 3.6 & 98 $\pm$ 25&  0.047 $\pm$ 0.005 &  83.5 $\pm$ 2.5 &  14 $\pm$ 1 &  2.5 $\pm$ 0.2 & 1.32 $\pm$ 0.13& 2&   [8] \\
4U 1254-690 \textbf{(S)}  & 51187.38&  1999-01-09  &  WFC  &   y  &  1 $\pm$ 0.3 &  13634 $\pm$ 4125&  0.30 $\pm$ 0.07 &   - &  > 14.9 $\pm$ 0.3 &  102 $\pm$ 4 & 349 $\pm$ 8&  - &  [9] \\
4U 1608-522 \textbf{(S)}  & 53495&   2005-05-05  &  ASM  &   y  &  3.8 $\pm$ 0.2 & 19348 $\pm$ 2379&  0.40 $\pm$ 0.09 &   -  &  > 28 $\pm$ 4 &  90 $\pm$ 10 & 309 $\pm$ 34&  - &  [10] \\
4U 1636-536 \textbf{(S)}  & 50253.61&   1996-06-19  &  ASM  &   -  &  1.75 $\pm$ 0.15 & 35856 $\pm$ 3354&  0.34 $\pm$ 0.08 &   - & > 17.5 $\pm$ 0.6 &  187 $\pm$ 7 & 641 $\pm$ 25&  - &  [11] \\
4U 1636-536 \textbf{(S)}  & 50642.37&   1997-07-13  &  ASM  &   -  &  1.5 $\pm$ 0.1 & 12304 $\pm$ 2383&  0.35 $\pm$ 0.08 &   - &  > 3.8 $\pm$ 0.7 &  55 $\pm$ 10 & 188 $\pm$ 34&  - &   [12] \\
4U 1636-536 \textbf{(S)}  & 51324.21&   1999-05-26  &  ASM  &   -  &  1.2 $\pm$ 0.1 & 17057 $\pm$ 2419&  0.38 $\pm$ 0.11 &   - &  > 6.1 $\pm$ 0.7 &  61 $\pm$ 7 & 209 $\pm$ 24&  - &   [13] \\
4U 1636-536 \textbf{(S)}  & 51962.7&   2001-02-22  &  PCA/ASM  &   y  &  2.2 $\pm$ 0.1 & 5034 $\pm$ 512&  0.18 $\pm$ 0.02 &   - &  84 $\pm$ 1 &  33 $\pm$ 3 & 113 $\pm$ 10&  - &  [11]; [14] \\
XTE J1701-407   & 54664.56&   2008-07-17  &   BAT/XRT  &   y  &  8 $\pm$ 2 & 80 $\pm$ 28&  0.011 $\pm$ 0.003 &  60 $\pm$ 1 &  > 5 $\pm$ 1.5 &  2.9 $\pm$ 0.7 & 1.51 $\pm$ 0.4&  - &   [1]; [15] \\
IGR J17062-6143  & 56103.94&   2012-06-25  &   BAT/XRT  &   y  &  4 $\pm$ 0.3 & 433 $\pm$ 86&   - &  80 $\pm$ 10 &  > 15 $\pm$ 3 &  11 $\pm$ 2 & 5.78 $\pm$ 1.05&  - &   [1]; [16] \\
IGR J17062-6143  & 57092.43&   2015-11-03  &   MAXI/XRT  &   -  &  6 $\pm$ 0.9 & 1179 $\pm$ 455&   - &  170 $\pm$ 10 &  > 72 $\pm$ 15 &  45 $\pm$ 16 & 23.66 $\pm$ 8.55&  - &   [17] \\
IGR J17062-6143  & 59022.04&   2020-06-22  &   MAXI/NICER  &   -  &  4.7 $\pm$ 0.6 & 278 $\pm$ 57&   - &   - & > 13 $\pm$ 2 &  8.3 $\pm$ 1.3 & 4.34 $\pm$ 0.72&  - &   [18]; [19] \\
4U 1705-44 \textbf{(S)}  & 57683&   2016-10-22  &   MAXI  &   -  &  7.1 $\pm$ 0.7 & 2784 $\pm$ 415&  0.77 $\pm$ 0.11 &   -  & > 12.4 $\pm$ 1.2 &  78 $\pm$ 8 & 268 $\pm$ 27&  - &   [5]; [20] \\
SAX J1712.6-3739  & 55830.84&   2011-09-26  &   BAT/XRT  &   y  &  8 $\pm$ 0.5 & 500 $\pm$ 140&   - &  160 $\pm$ 10 & > 1.6 $\pm$ 0.5 &  11 $\pm$ 3 & 5.78 $\pm$ 1.58&  - &   [1]; [21]; [22] \\
SAX J1712.6-3739  & 56887.71&   2014-08-18  &   BAT/XRT  &   y  &  6.9 $\pm$ 0.4 & 2580 $\pm$ 303&   - &  1000 $\pm$ 60 & > 5.4 $\pm$ 0.6 &  49 $\pm$ 5 & 25.63 $\pm$ 2.63&  - &   [1]; [21]; [23] \\
SAX J1712.6-3739  & 58169.96&   2018-02-20  &   JEM-X  &   y  &  8.32 $\pm$ 0.2 & 214 $\pm$ 15&   - &  30 $\pm$ 10 & 2.3 $\pm$ 0.2 &  4.9 $\pm$ 0.3 & 2.57 $\pm$ 0.13&  - &   [24] \\
*SAX J1712.6-3739  & 58246.38&   2018-05-08  &   BAT  &   - &   >10 $\pm$ 3 & 691 $\pm$ 276&    - &  210 $\pm$ 20 &  >70 $\pm$ 15 &  >19  & 9.86 $\pm$ 0&  - &   [1]; [21] \\
RX J1718-4029  & 50349.31&   1996-09-23  &   WFC  &   y &  4.6 $\pm$ 0.6 & 123 $\pm$ 19&  0.031 $\pm$ 0.015 &  91 $\pm$ 10 &  5.5 $\pm$ 0.5 &  2.5 $\pm$ 0.2 & 1.32 $\pm$ 0.13& 1718&   [25] \\
IGR J17254-3257  & 54009.3&   2006-10-01  &   JEM-X  &   y  &  1.5 $\pm$ 0.1 & 239 $\pm$ 82&   0.014 $\pm$ 0.002  &  60 $\pm$ 10 &  3.6 $\pm$ 1.2 &  9 $\pm$ 3 & 4.73 $\pm$ 1.32& 6229&  [24]; [26] \\
KS 1731-260 \textbf{(S)}  & 50349.42&   1996-09-23  &   WFC  &   y  &  2.1 $\pm$ 0.3 &  22552 $\pm$ 3480 &  0.11 $\pm$ 0.02 &   - & > 160 $\pm$ 9 &  120 $\pm$ 7 & 412 $\pm$ 25&  - &   [27] \\
Swift J1734.5-3027  & 56536.38&   2013-09-01   &   BAT/XRT  &   y &  8.8 $\pm$ 1.8 & 123 $\pm$ 28&  0.015 $\pm$ 0.008 &  100 $\pm$ 1 & > 10 $\pm$ 1.5 &  6.2 $\pm$ 1 & 3.22 $\pm$ 0.33&  - &  [1]; [28] \\
1RXH J173523.7-35401   & 54600.43&   2008-05-14  &   BAT/XRT  &   y  &  4.2 $\pm$ 0.3 & 354 $\pm$ 51&   0.0015 $\pm$ 0.0005  &  100 $\pm$ 50 & > 3.6 $\pm$ 0.2 &  16 $\pm$ 2 & 8.55 $\pm$ 1.32&  - &   [1]; [29] \\
SLX 1735-269  & 53108&   2004-04-13  &   ISGRI  &   y &  5.8 $\pm$ 0.2 & 257 $\pm$ 44&   -  &  320 $\pm$ 80 & > 15 $\pm$ 2 &  > 6  & 3.15 $\pm$ 0&  - &   [30] \\
SLX 1735-269  & 56267.15&   2012-12-06  &   MAXI  &   -  &  5.3 $\pm$ 0.7 & 350 $\pm$ 70&  0.14 $\pm$ 0.02 &   - & > 8 $\pm$ 1 &  45 $\pm$ 7 & 23.66 $\pm$ 3.95&  - &   [4]; [5], [31] \\
4U 1735-44 \textbf{(S)}  & 50318.13&  1996-08-23  &   WFC  &   -  &  2.1 $\pm$ 0.5 & 5075 $\pm$ 1433&  0.5  $\pm$ 0.23 &   - & > 65 $\pm$ 10 &  66 $\pm$ 10 & 226 $\pm$ 34&  - &   [32] \\
SLX 1737-282  & 53073.72&   2004-03-09  &   JEM-X  &   y  &  6.2 $\pm$ 0.1 & 573 $\pm$ 53&   0.012 $\pm$ 0.002 &  360 $\pm$ 25 &  5.4 $\pm$ 0.5 &  11 $\pm$ 1 & 5.78 $\pm$ 0.53& 4914&  [24]; [33] \\
SLX 1737-282  & 53471.34&   2005-04-11  &   JEM-X  &   y  &  5.7 $\pm$ 0.25 & 624 $\pm$ 63&   0.009 $\pm$ 0.002 &  360 $\pm$ 30 & > 5.4 $\pm$ 0.5 &  11 $\pm$ 1 & 5.78 $\pm$ 0.53& 5608&  [24]; [33] \\
SLX 1737-282  & 54192.24&   2007-04-02  &   JEM-X  &   y  &  5.1 $\pm$ 0.15 & 570 $\pm$ 66&   0.009 $\pm$ 0.002 &  230 $\pm$ 35 &  4.5 $\pm$ 0.5 &  9 $\pm$ 1 & 4.73 $\pm$ 0.53&  - &  [24]; [33] \\
XMM J174457-2850.3  & 56150.19&   2012-08-11  &   BAT/XRT  &   y  &  7.8 $\pm$ 0.5 & 119 $\pm$ 24&   -  &  50 $\pm$ 5 & > 0.25 $\pm$ 0.05 &  4.7 $\pm$ 0.9 & 2.43 $\pm$ 0.46&  - &   [34] \\
SAX J1747.0-2853  & 55605.52&   2011-02-13  &   JEM-X  &   -  &  5.6 $\pm$ 0.1 & 103 $\pm$ 15&  0.20 $\pm$ 0.02 &   -  & > 5.6 $\pm$ 0.6 &  1.4 $\pm$ 0.2 & 0.72 $\pm$ 0.07&  - &   [35] \\ 
SAX J1747.0-2853 \textbf{(S)}  & 55605.54&  2011-02-13  &  JEM-X/MAXI  &   y  &  3.4 $\pm$ 0.5 &  29074 $\pm$ 6463 &  0.20 $\pm$ 0.02 &   - & > 63 $\pm$ 8 &  240 $\pm$ 40 & 823 $\pm$ 137&  - &   [5]; [35] \\
SLX 1744-299  & 56388.46&   2013-04-06  &   JEM-X  &   y  &  4.3 $\pm$ 0.2 & 76 $\pm$ 7&   - &  14 $\pm$ 5 &  3.3 $\pm$ 0.2 &  2.8 $\pm$ 0.2 & 1.45 $\pm$ 0.13&  - &   [24] \\
SLX 1744-299  & 57080.41&   2015-02-27   &   JEM-X  &   (y)  &  3 $\pm$ 0.1 & 186 $\pm$ 24&   - &  30 $\pm$ 5 &  > 5.6 $\pm$ 0.7 &  4.8 $\pm$ 0.6 & 2.5 $\pm$ 0.33&  - &   [24] \\
GX 3+1 \textbf{(S)}  & 50973.04&  1998-06-09  &  ASM  &   -  &  2.5 $\pm$ 0.2 & 10550 $\pm$ 1528&   - &   - & > 8.3 $\pm$ 1.0 &  61.3 $\pm$ 7.4 & 210 $\pm$ 26&  - &   [5]; [36]\\
GX 3+1  & 53248.78&   2004-08-31  &   JEM-X  &   y  &  3.5 $\pm$ 0.3 & 603 $\pm$ 58&  0.18 $\pm$ 0.06 &   - &  21 $\pm$ 1 &  4.9 $\pm$ 0.2 & 2.57 $\pm$ 0.13& 5309&   [23]; [37] \\
EXO 1745-248 \textbf{(S)}  & 55858.53&  2011-10-24  &   MAXI  &   -  &  1.7 $\pm$ 0.3 & 56232 $\pm$ 11432&  0.09 $\pm$ 0.02 &  - & > 11 $\pm$ 1 &  63 $\pm$ 6 & 216 $\pm$ 21&  - & [5], [38], [39] \\
GRS 1747-312   & 52394.04&   2002-04-30  &   PCA  &   y &  2.9 $\pm$ 0.5 & 80 $\pm$ 15&   0.29 $\pm$ 0.04 &  69 $\pm$ 1 &  2.3 $\pm$ 0.1 &  1.24 $\pm$ 0.05 & 0.65 $\pm$ 0.03& 2994&   [40] \\
AX J1754.2-2754   & 59757.08&   2017-03-12  &   JEM-X  &   y  &  6.3 $\pm$ 0.55 & 85 $\pm$ 8&  0.008 $\pm$ 0.002 &  100 $\pm$ 10 &  7 $\pm$ 0.4 &  3.2 $\pm$ 0.1 & 1.71 $\pm$ 0.07&  - &   [24] \\
SAX J1806.5-2215  & 57844.79&   2017-04-01  &   BAT/XRT  &   y  &  4.2 $\pm$ 0.2 & 139 $\pm$ 36&   - &  130 $\pm$ 20 & > 4.4 $\pm$ 1.3 &  1.6 $\pm$ 0.4 & 0.85 $\pm$ 0.2&  - &   [1]; [41] \\
XTE J1810-189  & 55731.04&   2011-06-18  &   BAT/XRT  &   y  &  3.2 $\pm$ 0.4 & 121 $\pm$ 18&   - &  110 $\pm$ 20 & > 2.04 $\pm$ 0.04 &  1.5 $\pm$ 0.1 & 0.79 $\pm$ 0.07&  - &   [1] \\
GX 17+2 \textbf{(S)}  & 50340.3&  1996-09-14  &  WFC  &   -  &  1.3 $\pm$ 0.1 & 4566 $\pm$ 1051&  1.4 $\pm$ 0.1 &   - & > 52 $\pm$ 3 &  51.2 $\pm$ 11.1 & 175 $\pm$ 38&  - &   [42] \\
GX 17+2   & 50487.1&   1997-02-08  &   PCA  &   -  &  1.33 $\pm$ 0.01 & 245 $\pm$ 9&  2.2 $\pm$ 0.2 &  96 $\pm$ 16 & > 3.7 $\pm$ 0.1 &  2.8 $\pm$ 0.1 & 1.45 $\pm$ 0.07& 2261&   [43] \\
GX 17+2   & 51135.36&   1998-11-18  &   PCA  &   y  &  1.5 $\pm$ 0.12 & 279 $\pm$ 39&  1.8 $\pm$ 0.1 &  128 $\pm$ 16 &  4.7 $\pm$ 0.5 &  3.6 $\pm$ 0.4 & 1.9 $\pm$ 0.2& 2457&   [43] \\
GX 17+2 \textbf{(S)}  & 51444.1&  1999-09-23  &  WFC  &   -  &  1.7 $\pm$ 0.2 & 2462 $\pm$ 606&  0.9 $\pm$ 0.1 &   - & > 46 $\pm$ 5 &  36.1 $\pm$ 7.8 & 124 $\pm$ 0.19&  - &   [42] \\
GX 17+2 \textbf{(S)}  & 51452.33&  1999-10-01  &  WFC  &   y  &  2.2 $\pm$ 0.3 & 2345 $\pm$ 599&  1.2 $\pm$ 0.1 &   - & > 32 $\pm$ 4 &  44.5 $\pm$ 9.6 & 152 $\pm$ 33&  - &   [42] \\
GX 17+2   & 51454.65&   1999-10-03  &   PCA  &   y  &  1.32 $\pm$ 0.02 & 467 $\pm$ 8&  2.8 $\pm$ 0.3 &  136 $\pm$ 7 &  6.92 $\pm$ 0.02 &  5.31 $\pm$ 0.02 & 2.76 $\pm$ 0.01& 2582&   [43] \\
GX 17+2   & 51456.98&   1999-10-05  &   PCA  &   y  &  1.45 $\pm$ 0.06 & 90 $\pm$ 4&  2.8 $\pm$ 0.2 &  23 $\pm$ 1 &  1.46 $\pm$ 0.01 &  1.12 $\pm$ 0.01 & 0.59 $\pm$ 0.01& 2583&   [43] \\
GX 17+2   & 51457.46&   1999-10-06  &   PCA  &   y  &  1.41 $\pm$ 0.06 & 121 $\pm$ 6&  2.4 $\pm$ 0.15 &  44 $\pm$ 1 &  1.92 $\pm$ 0.01 &  1.47 $\pm$ 0.01 & 0.77 $\pm$ 0.01& 2584&   [43] \\
GX 17+2   & 51460.52&   1999-10-09  &   PCA  &   y  &  1.25 $\pm$ 0.03 & 120 $\pm$ 4&  2.4 $\pm$ 0.18 &  58 $\pm$ 2 &  1.68 $\pm$ 0.04 &  1.29 $\pm$ 0.03 & 0.68 $\pm$ 0.02& 2586&   [43] \\
GX 17+2   & 51461.38&   1999-10-10  &   PCA  &   y  &  1.45 $\pm$ 0.03 & 112 $\pm$ 17&  2.1 $\pm$ 0.1 &  24 $\pm$ 1 &  1.81 $\pm$ 0.27 &  1.4 $\pm$ 0.2 & 0.72 $\pm$ 0.13& 2587&   [43] \\
GX 17+2 \textbf{(S)}  & 51795.34&  2000-09-08  &  WFC  &   y  &  2.3 $\pm$ 0.4 & 958 $\pm$ 175&  1.9 $\pm$ 0.1 &   - & > 25 $\pm$ 1 &   > 19 $\pm$ 1 & 65 $\pm$ 3&  - &   [42] \\
GX 17+2   & 56011.77&   2012-03-25  &   JEM-X  &   y  &  5 $\pm$ 0.15 & 446 $\pm$ 18&  1.4 $\pm$ 0.2 &  260 $\pm$ 30 &  22 $\pm$ 5 &  19.2 $\pm$ 0.5 & 10.06 $\pm$ 0.27& 8706&   [24] \\
GX 17+2   & 56160&   2012-08-21  &   JEM-X  &   -  &  5.7 $\pm$ 0.2 & 369 $\pm$ 17&  0.9 $\pm$ 0.1 &  60 $\pm$ 10 & > 21 $\pm$ 5 &  18.1 $\pm$ 0.5 & 9.47 $\pm$ 0.27&  - &   [24] \\
4U 1820-30 \textbf{(S)}  & 51430&  1999-09-09  &  PCA  &   y  &  8.2 $\pm$ 0.3 & 3515 $\pm$ 375&   0.64 $\pm$ 0.08 &   - & > 200 $\pm$ 35 &  200 $\pm$ 20 & 685 $\pm$ 70&  - &   [44] \\
4U 1820-30 \textbf{(S)}  & 55272.72&  2010-03-17  &  MAXI  &   -  &  6.7 $\pm$ 0.6 & 1420 $\pm$ 470&  0.63 $\pm$ 0.11 &   - & > 82 $\pm$ 8 &  66 $\pm$ 21 & 226 $\pm$ 72&  - &   [4]; [5] \\
SAX J1828.5-1037 \textbf{(S)}  & 55877.34&  2011-11-12  &  MAXI  &   -  &  1.7 $\pm$ 0.4 & 5260 $\pm$ 1369&   - &   - & > 3.72 $\pm$ 0.22 &  36 $\pm$ 4 & 51 $\pm$ 14&  - & [4]; [5]; [45] \\
GS 1826-238 \textbf{(S)}  & 58161&  2018-02-18  &  MAXI  &   -  &  1.34 $\pm$ 0.21 & 14553 $\pm$ 2536&  0.23 $\pm$ 0.03 &   - & > 10.5 $\pm$ 0.2 &  105 $\pm$ 8 & 360 $\pm$ 27&  - &   [46] \\
Ser X-1 \textbf{(S)}  & 50507.08&  1997-02-28  &  WFC  &   -  &  2.3 $\pm$ 0.3 & 5183 $\pm$ 942&  0.34 $\pm$ 0.11 &   - & > 54 $\pm$ 7 &  87 $\pm$ 11 & 298 $\pm$ 38&  - &   [47] \\
Ser X-1 \textbf{(S)}   & 51399.14&  1999-08-09  &  ASM  &   -  &  1 $\pm$ 0.1 & 21511 $\pm$ 2415&   - &   - & > 7.9 $\pm$ 0.4 &  157 $\pm$ 8 & 538 $\pm$ 27&  - &   [12] \\
Ser X-1 \textbf{(S)}   & 54753.28&  2008-10-14  &  ASM  &   -  &  1.2 $\pm$ 0.1 & 5024 $\pm$ 477&   - &   - & > 43 $\pm$ 3 &  44 $\pm$ 2 & 151 $\pm$ 7&  - &   [12] \\
Ser X-1 \textbf{(S)}   & 55901.33 &  2011-12-06  &  MAXI  &   -  &  0.7 $\pm$ 0.3 & 9004 $\pm$ 4165&   - &   - & > 0.6 $\pm$ 0.1 &  46 $\pm$ 8 & 158 $\pm$ 27&  - &   [19] \\
4U 1850-086   & 56726 &   2014-03-10  &  BAT/XRT  &   y  &  12.1 $\pm$ 1 & 238 $\pm$ 36&   - &  380 $\pm$ 40 & > 25 $\pm$ 3 &  16.4 $\pm$ 2 & 8.61 $\pm$ 1.05&  - &   [4] \\
4U 1850-086   & 57151.39&   2015-05-09   &   MAXI  &   -  &  15 $\pm$ 1 & 418 $\pm$ 40&   - &   - & > 125 $\pm$ 8 &  35.7 $\pm$ 2.4 & 18.67 $\pm$ 1.25&  - &   [4] \\
Aql X-1 \textbf{(S)}  & 56493.3&   2013-07-20  &   MAXI  &   -  &  2.97 $\pm$ 0.46 & 23782 $\pm$ 4385&  0.75 $\pm$ 0.08 &   - & > 7.9 $\pm$ 0.8 &  150 $\pm$ 15 & 514 $\pm$ 51&  - &   [4]; [5] \\
Aql X-1 \textbf{(S)}  & 59130.7&   2020-10-08  &   MAXI  &   -  &  2.9 $\pm$ 0.3 & 12178 $\pm$ 3333&  - &  - & > 6 $\pm$ 2 &  60 $\pm$ 15 & 205 $\pm$ 50&  - &   [48] \\
M15 X-2  & 51871.59&   2000-11-23  &   WFC  &   y &  5 $\pm$ 1 & 247 $\pm$ 53&   0.12 $\pm$ 0.02 &  190 $\pm$ 11 & > 12.3 $\pm$ 0.9 &  6.4 $\pm$ 0.5 & 3.35 $\pm$ 0.27& 2028&  [49]; [50] \\

\label{tab:full}
\end{xltabular}

\normalfont
{\textbf{Ref}: [1] - \cite{intZ19}, [2] - \cite{defalco17a}, [3] - \cite{Kuul10}, [4] - \cite{Serino16}, [5] - \cite{intZ17b}, [6] - \cite{intZ05}, [7] - \cite{intZ11}, [8] - \cite{intZ08}, [9] - \cite{intZ03}, [10] - \cite{Keek08}, [11] - \cite{Wijnands01}, [12] - \cite{Kuul09atel}, [13] - \cite{Kuul04}, [14] - \cite{Stroh02}, [15] - \cite{linares09}, [16]  - \cite{degenaar13}, [17] - \cite{Keek17}, [18] - \cite{nashida}, [19] - \cite{Bult21}, [20] - \cite{iwak}, [21] - \cite{lin20},  [22] - \cite{palm11}, [23] - \cite{cummings14}, [24] - \cite{alizai}, [25] - \cite{Kaptein}, [26] - \cite{Chenevez07}, [27] - \cite{kuul02a}, [28] - \cite{bozzo15}, [29] - \cite{degenaar10}, [30] - \cite{sguera07}, [31] - \cite{negoro}, [32] - \cite{Cor00}, [33] - \cite{falan08}, [34] - \cite{degenaar14}, [35] - \cite{ATel3183}, [36] - \cite{kuul02c}, [37] - \cite{Chenevez06}, [38] - \cite{Serino12}, [39] - \cite{altam12}, [40] - \cite{intZ03a}, [41] - \cite{Barth17}, [42] - \cite{intZ04}, [43] - \cite{kuul02b}, [44] - \cite{Strohbrown02}, [45]- \cite{asada11}, [46] - \cite{iwak1}, [47] - \cite{Cor02}, [48] - \cite{iwak2}, [49] - \cite{intZ10}, [50] - \cite{intZ14b} \\\\
{\textbf\textbf{(S)} - Superburst. \\}
$^{1}$ $\times 10^{-8}$ \ergcs\\
$^{2}$ $\times 10^{-6}$ \ergcm\\
$^{3}$ $\times 10^{40}$ erg\\
$^{4}$ $10^{9}$ g cm$^{-2}$\\
$^{*}$ Adopted from (\cite{lin20})\\
The onset of the second burst from SLX~1744$-$299 is in parentheses as a significant rise in JEM-X counts was detected for 2~s before a slew of 3~min.
}
\end{landscape}

\twocolumn

\onecolumn
\begin{table}
\caption{
List of literature bursts and their parameters. $F_{\rm peak}$ is measured in a specific energy band of that particular instrument (see section \ref{subsection:literatur} for details). The $\tau$ values are lower limits as they are obtained from measurements of peak flux and total energy in a specific energy band of a particular instrument. None of the 14 bursts listed are categorized as superbursts.}
\begin{center}
\resizebox{\columnwidth}{!}{%
\begin{tabular}{l c c c c c c c}
\hline
\Centering
 {\textbf{Source}} & {\textbf{obs. date}} & {\textbf{Instrument}} & {\textbf{$F_{\rm peak}$ $^{1}$}} & {\textbf{$\tau$ (s) $^{¤}$}} & {\textbf{$E_{\rm tot}$ $^{2}$}} & {\textbf{ $y_{\rm b}$ $^{3}$}} & \textbf{Ref} \\
\hline

4U~0614$+$091 & 2002-02-17 & FREGATE &  $28.1 \pm 1.7$ & $216 \pm 12$ &  $4.88 \pm 0.1$ & $ 4.3 \pm 0.02$ & [1]\\
Cen~X$-$4 & 1969-07-07 & Vela 5B &  $140$ & $ \approx 145$ & $\approx 0.35 $ & $\approx 0.23$ & [2]\\
4U~1708$-$23 & 1979-02-07 & SAS-3 &  $\approx 2.5$ & $ \approx 308$ &  $\approx 9.2 $ & $\approx 6$ & [3]\\
3A~1715$-$321 & 1982-07-20 & Hakucho &  $\approx 6.7$ & $ \approx 336$ & $\approx 8.0 $ & $\approx 5.4$ & [4];[5]\\
4U~1722$-$30 & 1979-03-05 & Einstein &  $\approx 4.17$ & $ \approx 55$ & $\approx 1.5$ & $\approx 1$ & [6]\\
SLX~1735$-$269 & 2003-09-15 & JEM-X & $20 \pm 5$ & $161 \pm 18$ & $13 \pm 3$ & $32 \pm 6$ & [7]\\
SLX~1735$-$269 & 2005-06-20 & FREGATE/WXM & N/a & $<400$ & N/a & N/a & [8]\\
GX~17$+$2 & 1981-05-25 & Hakucho &  $>0.54$ & $253 \pm 44$ & $>1.18$ & $>0.77$ & [9]\\
GX~17$+$2 & 1981-05-25 & Hakucho &  $>0.71$ & $88 \pm 18$ & $>0.54$ & $>0.35$ & [9]\\
GX~17$+$2 & 1982-08-13 & Hakucho &  $>0.59$ & $53 \pm 11$ & $>0.27$ & $>0.18$ & [9]\\
GX~17$+$2 & 1982-08-16 & Hakucho &  $>0.74$ & $270 \pm 52$ & $>1.73$ & $ >1.1$ & [9]\\
GX~17$+$2 & 1985-08-20 & EXOSAT &  $>1.4$ & $262$ & $>3.17$ & $ >2.1$ & [10]\\
SLX~1744$-$299 & 1990-10-09 & Granat &  $\approx 3.5$ & $ \approx 60$ & $\approx 1.8 $ & $\approx 1.2$ & [11]\\
M15~X$-$2 & 1988-08-20 & Ginga/LAC &  $\approx 3.75$ & $287$ & $\approx 5.6$ & $\approx 3.6$ & [12]\\

\hline
\end{tabular}%
}
\end{center}

Ref: {[}1{]} - \cite{Kuul10}, {[}2{]} - \cite{Kuul09}, {[}3{]} - \cite{Lew84},  {[}4{]} - \cite{tawaraa}, {[}4{]} - \cite{Kuul10}, {[}5{]} - \cite{tawarab}, {[}6{]} - \cite{Grindlay80}, {[}7{]} - \cite{molkov05}, {[}7{]} - \cite{Lew84}, {[}8{]} - \cite{suzuki05}, {[}8{]} - \cite{tawaraa}, {[}9{]} - \cite{tawarac}, {[}10{]} - \cite{sztajno}, {[}11{]} - \cite{Pav}, {[}11{]} - \cite{molkov05}, {[}12{]} - \cite{vanParadijs90} \label{tab:lit}
\\
$^{1}$ $\times 10^{-8}$ \ergcs\\
$^{2}$ $\times 10^{40}$ ergs\\
$^{3}$ $10^{9}$ g cm$^{-2}$\\
\label{tab:lit} 
\end{table}
\twocolumn


\begin{table}
\caption{18 X-ray bursts that were too short to be included in our catalogue of long bursts, but with significant longer duration and/or longer PRE-phase than the classical X-ray bursts.}
\resizebox{\columnwidth}{!}{%
\begin{tabular}{c|c|c|c|c|c}
\hline
\centering
 Source    & MJD & Inst. &  $\tau$(s) & $E_{\rm tot}$\textbf{\textsuperscript{\textdagger}} & Ref.\\ \hline
4U 0513-40 & 50881.50 &  WFC & 38 $\pm$ 15 &  1.47 $\pm$ 0.21& [1]\\
4U 0513-40 & 52142.16 &  WFC & 38 $\pm$ 15 &  1.24 $\pm$ 0.11& [1]\\
4U 0614+091 &  51944.91 & HETE-2 & $39.7 \pm 2$  & 2.55 $\pm$ 0.15& [2]\\
2S 0918-549 & 51339.05 & WFC & $36 \pm 4$ & 1.16 $\pm$ 0.05 & [3]\\
4U 1246-588 & 50286.29 & WFC & 60.9 $\pm$ 10.1 & 2.37 $\pm$ 0.05 & [4]\\
4U 1246-588 & 51539.87 & WFC & 31.1 $\pm$ 6  & 1.02 $\pm$ 0.05 & [4]\\
4U 1246-588 & 51929.87 & WFC & 41.4 $\pm$ 10.5  & 1.54 $\pm$ 0.07 & [4]\\
4U 1246-588 & 53958.12 & BAT/XRT & 43 $\pm$ 12  & 1.10 $\pm$ 0.3 & [4]\\
XTE J1701-407 & 54674.93 & BAT/XRT & 60 $\pm$ 20  & 1.10 $\pm$ 0.40 & [5]\\
SAX J1712.6-3739  & 55378.62 & BAT/XRT & 64 $\pm$ 22 & 0.9 $\pm$ 0.3 & [6]\\
4U 1722-30  & 50395.29 & PCA & 33 $\pm$ 4  & 1.4 $\pm$ 0.2 & [7] \\
4U 1722-30  & 54526.68 & PCA & 27 $\pm$ 3  & 0.52 $\pm$ 0.06 & [8]\\
GRS 1741.9-2853 & 56507.96 & NuSTAR & 45 $\pm$ 8 & 0.97 $\pm$ 0.10 & [9]\\
SLX 1744-299  & 50325.05 & WFC & 32.9 $\pm$ 4.5 & 1.63 $\pm$ 0.09 & [10]\\
SLX 1744-299  & 50367.03 & WFC & 33 $\pm$ 7 & 0.96 $\pm$ 0.07 & [10]\\
SLX 1744-299  & 57454.43 & JEM-X & 56 $\pm$ 9 & 1.9 $\pm$ 0.3 & [11]\\
AX J1754.2-2754 & 53476.92 & JEM-X & 35 $\pm$ 3 & 1.7 $\pm$ 0.1 & [11]\\
IGR J18245-2452 & 56381.63 & BAT/XRT & 28 $\pm$ 10 & > 0.3 & [12]\\

\hline
\end{tabular}%
}

    \normalfont
    {\textbf{\textsuperscript{\textdagger}} $\times$ 10$^{40}$ erg.\\
    \textbf{Ref}: {[}1{]} - \cite{Gal20}, {[}2{]} - \cite{Kuul10}, {[}3{]} - \cite{intZ05}, {[}4{]} - \cite{intZ08}, {[}5{]} - \cite{linares09}, {[}6{]} - \cite{lin20}, {[}7{]} - \cite{molkov00},  {[}8{]} - This work,  {[}9{]} - \cite{barr13}, {[}10{]} - \cite{intZ07}, {[}11{]} - \cite{alizai}, {[}12{]} - \cite{Barth13}
    }
    \label{tab:short_bursts}
\end{table}

 All bursts from GX~17$+$2 are considered separately from other intermediate-duration bursts and superbursts in the remainder of this paper. At the time of publication of this study, no consistent explanation exists that explains all the different types of X-ray bursts observed from GX~17$+$2. Further long-term monitoring is needed to understand the nature of the source itself. 
 
\section{Summary and Conclusion}
\label{sec:conclusion}
Based on the criteria defined in \ref{subsec:dataselec}, we have collected a complete catalogue of 84 rare long bursts that most likely do not involve the rp-process burning.
We have performed a systematic reanalysis of archival data from seven different observatories, allowing us to provide light curves and time-resolved spectroscopic analyses for 70 bursts. Based on a literature study we include 14 more long bursts observed with the early generation of X-ray instruments.  

We conclude that intermediate-duration bursts occur at accretion rates of $0.001-0.03\dot{M}_{\rm Edd}$, with some notable outliers (GRS~1747$+$312 and GX~3$+$1) almost up to $0.3\dot{M}_{\rm Edd}$. The majority of superbursts occur at accretion rates of $\gtrsim 0.3\dot{M}_{\rm Edd}$ with a couple outliers being the superbursts from 4U~0614$+$091 with an accretion rate of $0.01\dot{M}_{\rm Edd}$.
It also appears from our study that values of $\tau$ and \textit{$E_{\rm tot}$} are the main observational discriminators between intermediate-duration bursts and superbursts. Indeed, the timescale of intermediate-duration bursts seems to remain below $\tau=1000$~s. They also release at least $10^{41}$~erg, but at most $3\times 10^{41}$~erg, above which energy we only find superbursts.
The $y_{\rm b}$ parameter, that is directly related to the total energy release, gives an indication of how deep a particular burst has ignited in the neutron star envelope. This is again a confirmation that the two types of long bursts arise from two different burning locations and thus from likely different burning fuels.

We include the long bursts from GX~17$+$2 in our catalogue but do not discuss them in the context of the other long bursts. Both intermediate-duration bursts and superbursts from GX~17$+$2 occur at very high accretion rates, indicating high enough temperatures for Helium to be burnt stably into Carbon that eventually get destroyed in further reactions. Bursts from this source are most likely Hydrogen-rich, making them different than the other long bursts.    

We find a group of long bursts originating from slow accretors (predominantly UCXBs) that bridges the gap between classical bursts and intermediate-duration bursts. These bursts have significantly higher energy release than the classical bursts (constituting $99\%$ of all X-ray bursts).
Previous studies have indicated that sources showing bursts with a continuous distribution of durations and energies may have a He poor white dwarf as the donor. This may also be the case for the 11 sources from which the 18 marginally-long bursts originate, as seven of them are categorized as UCXBs and four as transients. 

Our observational work provides the largest database of long X-ray bursts. Light curves and spectroscopic results can be downloaded from \href{http://doi.org/10.11583/DTU.21914916}{doi:10.11583/DTU.21914916}.  
The natural continuation of the present work would be to fit the bolometric light curves of the bursts provided here with cooling models, systematically investigating the burst energetics and ignition depths.
Furthermore, it will be interesting to investigate further the
marginally-long bursts introduced in our study that seem to bridge a gap.


\section*{Acknowledgements}
This work is based on the long-term observations performed by
the INTEGRAL international astrophysical gamma-ray observatory
and retrieved via the Russian and European INTEGRAL Science
Data Centers.
Quick-look results are provided by the ASM/RXTE team.This team includes all those working on the ASM at MIT and at the Goddard Space Flight Center SOF and GOF. 
This research has made use of data obtained through the High Energy Astrophysics Science Archive Research Center Online Service, provided by the NASA/Goddard Space Flight Center.
This work made use of data supplied by the UK Swift Science Data Centre at the University of Leicester.
ND is supported by a Vidi grant from the Netherlands Organisation for Scientific Research (NWO).
Our sincere thanks to Hauke Worpel for the inclination ranges used in this study and to Søren Brandt and Niels Lund for the fruitful discussions and feedback.
We thank the International Space Science Institute (ISSI) and the original ISSI team on thermonuclear bursts. The former for their essential financial support and the latter for assembling the first version of long burst catalogue.
Lastly, the authors thank the anonymous referee whose comments have helped improve this paper significantly.

\section*{Data Availability}
Along with the present paper, the catalogue, including light curves and time-resolved spectroscopy, is available for online download (\href{http://doi.org/10.11583/DTU.21914916}{http://doi.org/10.11583/DTU.21914916}).

\end{document}